\documentclass[a4paper,11pt]{article}

\usepackage{amsmath}
\usepackage{amsfonts}
\usepackage{amssymb}
\usepackage{graphicx}
\usepackage{subfig}   

\newcommand{\p}{\ensuremath{\partial}}
\newcommand{\del}{\ensuremath{\delta}}
\newcommand{\Del}{\ensuremath{\Delta}}
\newcommand{\eps}{\ensuremath{\varepsilon}}
\newcommand{\ep}{\ensuremath{\epsilon}}

\newcommand{\Om}{\ensuremath{\Omega}}

\newcommand{\sig}{\ensuremath{\sigma}}

\newcommand{\Msol}{\ensuremath{M_{\rm sol}}}
\newcommand{\etal}{\emph{et al}.}

\renewcommand{\k}{\ensuremath{\mathbf{k}}}
\newcommand{\x}{\ensuremath{\mathbf{x}}}
\newcommand{\tW}{\ensuremath{\widetilde{W}}}
\newcommand{\R}{\ensuremath{\mathcal{R}}}

\newcommand{\Cite}[1]{Ref.~\cite{#1}}
\newcommand{\Cites}[1]{Refs.~\cite{#1}}
\newcommand{\eqn}[1]{Eqn.~\eqref{#1}}
\newcommand{\eqns}[1]{Eqns.~\eqref{#1}}
\newcommand{\fig}[1]{Fig.~\ref{#1}}
\newcommand{\figs}[1]{Figs.~\ref{#1}}
\newcommand{\ph}[1]{\phantom{#1}}

\newcommand{\be}{\begin{equation}}
\newcommand{\ee}{\end{equation}}
\newcommand{\la}[1]{\label{#1}}

\newcommand{\Cal}[1]{\ensuremath{\mathcal{#1}}}
\newcommand{\ti}[1]{{\tilde #1}}
\newcommand{\avg}[1]{\ensuremath{\langle \,#1\, \rangle}}
\newcommand{\delc}{\ensuremath{\delta_{\rm c}}}
\newcommand{\delv}{\ensuremath{\delta_{\rm v}}}
\newcommand{\delT}{\ensuremath{\delta_{\rm T}}}
\newcommand{\nuc}{\ensuremath{\nu_{\rm c}}}
\newcommand{\nuv}{\ensuremath{\nu_{\rm v}}}
\newcommand{\nuT}{\ensuremath{\nu_{\rm T}}}
\newcommand{\DS}{\ensuremath{\Delta S}}
\newcommand{\fnl}{\ensuremath{f_{\rm NL}}}
\newcommand{\gnl}{\ensuremath{g_{\rm NL}}}

\newcommand{\Fsw}{\ensuremath{\Cal{F}_{\rm SvdW}}}
\newcommand{\fsw}{\ensuremath{f_{\rm SvdW}}}
\newcommand{\lsw}{\ensuremath{\Cal{L}_{\rm SvdW}}}

\newcommand{\Fng}{\ensuremath{\Cal{F}_{\rm NG}}}
\newcommand{\fng}{\ensuremath{f_{\rm NG}}}
\newcommand{\Fws}{\ensuremath{\Cal{F}_{\rm WS}}}

\begin{document}

\begin{center}
\Large{\textbf{Excursion Sets and Non-Gaussian Void Statistics}} \\[0.5cm]
 
\large{Guido D'Amico$^{\rm a,d,e}$, Marcello Musso$^{\rm b}$,
Jorge Nore\~na$^{\rm c,d,e}$, Aseem Paranjape$^{\rm b}$}
\\[0.5cm]

{\renewcommand{\thefootnote}{}\footnotetext[0]{E-mail:
    gda2@nyu.edu, musso@ictp.it, jorge.norena@icc.ub.edu,
    aparanja@ictp.it}}

\small{
\textit{$^{\rm a}$ Center for Cosmology and Particle Physics,\\
Department of Physics, New York University,\\
4 Washington Place, New York, NY 10003, USA}}

\vspace{.2cm}

\small{
\textit{$^{\rm b}$ Abdus Salam International Centre for Theoretical
  Physics\\ Strada Costiera 11, 34151, Trieste, Italy}} 

\vspace{.2cm}

\small{
\textit{$^{\rm c}$ 
Institut de Ci\`encies del Cosmos (ICC),\\
Universitat de Barcelona (IEEC-UB), Mart\'i Franqu\`es 1, E08028
Barcelona, Spain}}

\vspace{.2cm}

\small{
\textit{$^{\rm d}$ SISSA, via Bonomea 265, 34136 Trieste, Italy}}

\vspace{.2cm}

\small{
\textit{$^{\rm e}$ INFN - Sezione di Trieste, 34151 Trieste, Italy}}

\end{center}

\vspace{.8cm}

\hrule \vspace{0.3cm}
\noindent \small{\textbf{Abstract}\\
Primordial non-Gaussianity (NG) affects the large scale structure
(LSS) of the universe by leaving an imprint on the distribution of
matter at late times. Much attention has been focused on using the
distribution of collapsed objects (i.e. dark matter halos and the 
galaxies and galaxy clusters that reside in them) to probe primordial
NG. An equally interesting and complementary probe however is the
abundance of extended \emph{underdense} regions or \emph{voids} in the
LSS. The calculation of the abundance of voids using the excursion set
formalism in the presence of primordial NG is subject to the same
technical issues as the one for halos, which were discussed e.g. in
\Cite{D'Amico:2010ta}. However, unlike the excursion set problem for
halos which involved random walks in the presence of one barrier
\delc, the void excursion set problem involves \emph{two} barriers
\delv\ and \delc. This leads to a new complication introduced by what
is called the ``void-in-cloud'' effect discussed in the literature,
which is unique to the case of voids. We explore a path integral
approach which allows us to carefully account for all these
issues, leading to a rigorous derivation of the effects of primordial
NG on void abundances. The void-in-cloud issue in particular makes the
calculation conceptually rather different from the one for halos.
However, we show that its final effect can be described by a simple yet
accurate approximation. Our final void abundance function is valid on
larger scales than the expressions of other authors, while being
broadly in agreement with those expressions on smaller scales.\\  
\noindent
\hrule
\vspace{0.5cm}
}
\setcounter{footnote}{0}

\section{Introduction}
\la{intro}
\noindent
A striking feature of the large scale structure (LSS)
of the universe which has emerged from studies over the last few
decades is the presence of a filamentary network or cosmic web
in the matter distribution, with galaxies distributed along filaments
which surround large apparently empty regions termed \emph{voids} 
\cite{Kirshner:1981wz,Patiri:2005ys,Goldberg:2004jr,
  Croton:2004ac,Hoyle:2001kn,Hoyle:2005kn}. A considerable amount of
analytical and numerical effort has gone into understanding the nature
of the cosmic web (see e.g. \Cites{ForeroRomero:2008ig,
  Lee:2008ar,AragonCalvo:2007mk,Shandarin:2009ue}, for a review see
\Cite{vandeWeygaert:2009vy}). Apart from being intrinsically
interesting, the LSS has the potential to be a powerful probe of
cosmology since it is sensitive to both the expansion history as well
as the initial conditions (in particular the physics of
inflation). The advent of large galaxy surveys has realised this
potential considerably over the last decade or so
\cite{Peacock:2001gs,Eisenstein:2005su,Abazajian:2008wr}, with
ongoing and upcoming surveys set to significantly increase the
precision of the LSS as a cosmological tool
(see e.g. \Cites{Cimatti:2009is,Cappelluti:2010ay,Schlegel:2009hj}).

Our focus in this work is on probing the physics of the
early universe. In the current understanding of structure formation,
the present LSS has its seeds in the statistics of the tiny primordial
curvature inhomogeneities generated in the very early 
universe, during the rapidly expanding inflationary phase. The
simplest models of inflation involving a single ``slowly rolling''
scalar field predict that the statistics of these primordial
inhomogeneities are almost Gaussian
\cite{Maldacena:2002vr,Acquaviva:2002ud}. Constraining 
the amount of \emph{non}-Gaussianity (NG) in the primordial
distribution therefore provides a unique window into the physics of
inflation. Until recently, the cosmic microwave background (CMB)
radiation was considered the standard tool for constraining primordial
NG, since inhomogeneities at the CMB epoch are small
and the physics can be described by a perturbative treatment. In terms
of the standard parametrization of the NG, the CMB constraints for the
local model translate to $-10 < f^{loc}_{NL} < +74$
\cite{Komatsu:2010fb}. (For a brief introduction to primordial NG see
Appendix \ref{app-primNG}.) Over the last several years, however, the
LSS has been shown to be an equally constraining and moreover
complementary probe of primordial NG. For example, from
\Cite{Slosar:2008hx} one finds $-29 < f^{loc}_{NL} < +69$, already
comparable with the CMB constraints, with precisions of order $\Delta
f^{loc}_{NL} \sim 10$~\cite{Sartoris:2010cr} and $\Delta f^{loc}_{NL}
\sim 1$~\cite{Cunha:2010zz} being claimed for future surveys.  
These constraints and forecasts rely on the statistics
of massive collapsed dark matter halos and the galaxies and galaxy
clusters that reside in them, chiefly using three tools -- the scale
dependent bias in the galaxy power spectrum \cite{Dalal:2007cu,
  Matarrese:2008nc}, the galaxy bispectrum \cite{Sefusatti:2009qh} and
the number density of collapsed objects (mass function)
\cite{Matarrese:2000iz,LoVerde:2007ri}. For recent reviews, see
\Cites{Verde:2010wp, Desjacques:2010jw}. 

On the other hand, the voids found in the LSS are also potentially
interesting probes of NG. While any given galaxy survey is expected to
contain fewer voids than halos, 
the effect of NG on void abundances is opposite to that on halo
abundances (see below). Whereas a positive \fnl\ will enhance the
abundance of halos, it will reduce the abundance of voids and vice
versa, making voids a complementary probe of NG. Although the
literature contains several definitions of what a void exactly means (see
e.g. \Cite{Colberg:2008qg} for a recent review and thorough
comparison), the basic physical picture is that of a large expanding
region which is underdense compared to the background. For a review of
the structure and dynamics of cosmic voids, see
\Cite{vandeWeygaert:2009hr}. From the point of view of using voids as
statistical probes, detailed numerical studies
\cite{Dubinski:1992tr,Colberg:2004nd,vandeWeygaert:1993cp}
and analyses \cite{Blumenthal:1992tr,Sheth:2003py} seem to indicate
that the excursion set formalism
\cite{Press:1973iz,Epstein:1983cp,Bond:1990iw}, combined with
simplified analytical models which track the underdensity of dark 
matter in a region, provides a good starting point. More precisely,
the spherical ansatz
\cite{Gunn:1972sv,Fillmore:1984nj,Bertschinger:1985nj,Blumenthal:1992tr}  
which we discuss later in a bit more detail has been found to be a
very useful tool when studying voids.  (See also
\Cites{Lavaux:2009wm,Biswas:2010ey} for a different approach and
application.) 

As one might expect from experience with halo abundances, analytical
treatments of void abundances are subject to some caveats. Firstly, by
choosing to describe voids using underdensities in the \emph{dark
  matter} distribution, one is a step removed from actual observations
which involve \emph{galaxies}. As pointed out by Furlanetto \& Piran
\cite{Furlanetto:2005cc}, the visually striking empty patches of
galaxy surveys are all \emph{galaxy} voids, and it requires some
nontrivial analysis to relate their distribution to that of the
underlying dark matter. There is also some indication from 
$N$-body simulations (with Gaussian initial conditions)
\cite{Colberg:2004nd} that analyses of the distribution of voids based 
on the excursion set formalism, such as the one by Sheth \& van de
Weygaert \cite{Sheth:2003py}, could fail to capture some non-universal
features of the void abundance. Another concern is that the
spherical assumption is of course an idealization -- real structure is
more complicated. Nevertheless, from the point of view of obtaining a
better understanding of the underlying physics from an analytical
perspective, an excursion set analysis based on the spherical ansatz
remains the most robust starting point currently available. Once the
calculation in this simplified setting is under control, one can
explore improvements by relaxing the assumptions involved. We will
adopt this point of view in this work, our goal being to obtain an
expression for the abundance of voids in the presence of  primordial
NG. More precisely, we are after the differential comoving number
density $dn_{\rm com}/dR_{\rm com}$ of voids of comoving size $R_{\rm
  com}$ which satisfies  
\be
R_{\rm com}\frac{dn_{\rm com}}{dR_{\rm com}} = \frac12\, \frac{3}{4\pi 
  R^3}\,f(S)\, \left| \frac{d\ln S}{d\ln R} \right|\,, 
\la{numbercounts}
\ee
where $R$ is the \emph{Lagrangian} radius of the void related to the
comoving radius by $R_{\rm com}=1.7R$ (see below), 
$S\equiv\sig^2(R)$ is the variance of the density contrast smoothed on
scale $R$, and the multiplicity function $f(S)$ is what we will
calculate analytically based on the excursion set formalism. Our
results will be in a form that should be testable in $N$-body
simulations which have complete control on the dark matter 
distribution. We will leave the second (and important) part of the
exercise, namely that of relating our results to observationally
relevant quantities, to future work. 

We should mention that this approach has been adopted by other
authors before us \cite{Kamionkowski:2008sr,Lam:2009nd} (see also
\Cite{Chongchitnan:2010xz}). The analysis of Kamionkowski
\etal\ \cite{Kamionkowski:2008sr}, while pioneering, was based on
several simplifying assumptions which we believe give an 
incomplete picture of the problem. In particular they used a
Press-Schechter-like approach for the statistics which can be shown to
miss certain scale dependent terms in the multiplicity $f$ that arise
from somewhat complex multi-scale correlations
\cite{Maggiore:2009rx,D'Amico:2010ta}. Additionally their treatment of
the non-Gaussianities involved an Edgeworth-like expansion (based on
\Cite{LoVerde:2007ri}) for the one point probability density of the
smoothed density contrast, linearized in the non-Gaussianity
amplitude. As pointed out in \Cite{D'Amico:2010ta} 
for the case of dark matter halos, such an expansion tends to
underestimate the effect of the NG on large scales, which is the
regime we wish to probe using voids. And finally, Kamionkowski
\etal\ also ignored a complication which arises when studying voids,
which Sheth \& van de Weygaert termed the ``void-in-cloud'' issue
(which we discuss in detail below). Lam \etal\ \cite{Lam:2009nd} on
the other hand used a more rigorous approach which accounted for
multi-scale correlations and the void-in-cloud problem, but their
treatment was also based on a linearized Edgeworth-like expansion and
is hence subject to the same caveat mentioned above. Additionally
their analysis made a technical assumption regarding the multi-scale
correlations which is not strictly valid (see below). Our treatment
below will be based on techniques introduced by Maggiore \& Riotto 
\cite{Maggiore:2009rv,Maggiore:2009rw} and developed further by
D'Amico \etal\ \cite{D'Amico:2010ta}. This will firstly allow us to
derive a void multiplicity function which is valid on larger scales
than the expressions of other authors. Secondly our approach will
allow us to carefully account for the void-in-cloud issue (without
making assumptions regarding the multi-scale correlations), and we
will show that its final effects can actually be described using a
simple yet accurate approximation.

The plan of this paper is  as follows. In Section \ref{SvdWbasic} we
review the excursion set analysis of Sheth \& van de Weygaert based
on the spherical ansatz for Gaussian initial conditions, and describe
the void-in-cloud issue mentioned above. To carefully account for this
problem, we turn to a path integral description. We begin in Section
\ref{SvdWpathinteg} by rederiving the Gaussian result for the void
multiplicity using path integrals, which allows us to introduce some 
formal machinery as well as discuss some of the subtleties of
the calculation in a controlled setting. Following this, in Section
\ref{NGvoids} we generalize the calculation to the non-Gaussian case
and derive our main result for the non-Gaussian void multiplicity,
showing how to account for the void-in-cloud issue. Our final result
is given in \eqn{finalresult} and is plotted in \figs{fig-ratio} and
\ref{fig-mult}. We conclude in Section \ref{discuss} with a brief
discussion of the result and prospects for future work. Several
technical details have been relegated to the Appendices.   

\section{Spherical void statistics: the SvdW result}
\la{SvdWbasic}
\noindent
The spherical ansatz for an expanding underdense region allows for a
completely analytical treatment, exactly like what happens in the
usual spherical \emph{collapse} model \cite{Gunn:1972sv}. Detailed 
numerical treatments \cite{Hoffman:1984nj,Dubinski:1992tr}
also indicate that, although simplistic, this ansatz goes a long way in
describing the evolution of individual underdense regions in realistic
settings. For mathematical details of the model we refer the reader to
\Cite{Sheth:2003py}, where Sheth and van de Weygaert (henceforth SvdW) 
also extensively discussed the physics of voids in the context of
hierarchical structure formation. Since our primary concern in this
paper is the statistical model built using the spherical ansatz, we
will restrict ourselves to briefly describing some of the physical
aspects relevant to our treatment. 

An important feature of the spherical expansion scenario is the
phenomenon of \emph{shell crossing}
\cite{Blumenthal:1992tr,Sheth:2003py}. Physically shell crossing 
occurs because inner mass shells decelerate slower than the outer
ones. With a sufficiently steep initial profile, this difference in
decelerations is large enough that the inner shells catch up with the
outer shells in a finite time, piling up mass in a very sharp
ridge-like feature which then evolves self-similarly
\cite{Bertschinger:1985nj}. It can then be argued that the time of the
\emph{first} shell crossing is a sensible definition of the time of
void formation. Such a definition brings us square into the realm of 
barriers and random walks, with the linearly extrapolated overdensity 
\delc\ from spherical collapse being replaced by a linearly
extrapolated underdensity $(-\delv)$ corresponding to the time of
first shell crossing. For an Einstein-de Sitter cosmology and assuming
a tophat initial profile for the underdensity, this shell crossing
threshold evaluates to $\delv=2.72$
\cite{Blumenthal:1992tr}. Note that the excursion set ansatz 
relies on smoothing the (linearly extrapolated) initial conditions
with a \emph{Lagrangian} scale $R$ associated with a conserved mass
$M\propto R^3$. The observationally relevant scale however is the
\emph{comoving} scale $R_{\rm com}$ which corresponds to the physical
size today of a sphere reaching first shell crossing. The spherical
model predicts the relation \cite{Blumenthal:1992tr}
\be
R_{\rm com} = 1.7 R\,.
\la{com-vs-Lag}
\ee
%
It might seem that the statistical problem involving voids is now
identical to the one with collapsed halos, and that a simple
replacement $\delc\to-\delv$ in all halo statistics results would
suffice to give the corresponding void statistics. The situation is
not this simple however, as SvdW discuss extensively in
\Cite{Sheth:2003py}. Recall that the collapse problem had to deal with
the subtlety of the ``cloud-in-cloud'' issue. This was the fact that,
when counting the fraction of collapsed objects of mass $M$, one must
only consider those trajectories (of the random walk of the smoothed
density contrast) which \emph{first} crossed the threshold \delc\ at a
smoothing scale $M$, as the smoothing radius is
decreased. Physically this corresponds to excluding overdense regions
which are embedded in bigger overdense regions, since in such a case
only the bigger region would survive as an independent collapsed
object. Due to the nature of the problem, of course a similar
requirement also holds in the case of voids, which SvdW call the
``void-in-void'' issue. This can be dealt with exactly as in the
collapse case, and requires us to consider only first-crossing
scales. 

Additionally, SvdW also identified a second problem which is unique to
the case of voids. Physically, this is the scenario where a region of
size $R_1$ satisfies the threshold underdensity requirement and is a
potential void candidate, but happens to be embedded in an
\emph{overdense} region of size $R_2>R_1$ that satisfies the collapse
criterion. This bigger region will then form a collapsed object,
crushing the underdense void candidate out of existence. Explicit
examples of such cases in $N$-body simulations were shown in
\Cite{Sheth:2003py}. The problem is therefore to exclude such
situations from the statistics and was labeled the ``void-in-cloud''
issue by SvdW, who also showed how the problem could be tackled. The
basic idea is that for the void formation problem there are now
\emph{two} barriers which are relevant -- the negative void shell
crossing threshold $(-\delv)$ which we will call the ``void barrier'',
and also the positive halo formation threshold \delc\ which we will
call the ``halo barrier''\footnote{The actual value of \delc\ must be
  chosen carefully, since simply setting it equal to the usual
  spherical collapse value of $1.686$ physically amounts to only
  excluding those underdense regions that would be completely crushed
  by collapsing overdensities. As SvdW discuss, this all-or-nothing
  approach fails to account for intermediate underdensities which are
  in the process of being crushed, thereby overestimating e.g. the
  typical void size in the Gaussian case (see also
  \Cite{Furlanetto:2005cc}). In this work we will ignore this
  complication and assume $\delc=1.686$.}.  
Statistically, excluding the void-in-cloud cases amounts to
counting only those trajectories in which the void barrier is first
crossed \emph{before} the halo barrier is ever crossed, as the
smoothing radius is decreased from large values. SvdW showed that
accounting for the void-in-cloud issue qualitatively changes the
behaviour of the void multiplicity as compared to the halo
multiplicity, by introducing a cutoff at small smoothing radii. This
is again intuitively clear since small underdense regions would be
more likely to find themselves embedded in larger scale overdensities,
and should therefore not survive as voids. Our interest when extending
the problem to the case of non-Gaussian initial conditions will
primarily be in the \emph{large} radius end, which is where
non-Gaussianities are expected to play a significant role.  

Before proceeding, a word on notation.  In addition to the void
barrier $(-\delv)$ and the halo barrier \delc, we will frequently
also require the \emph{total} barrier height
\be
\delT\equiv\delc+\delv\,. 
\la{delT-def}
\ee
Additionally, all our later expressions will involve the ``dimensionless''
parameters $\del/S^{1/2}$, where $S=S(R) = \avg{\del_R^2}$ is the
variance of the density contrast smoothed on scale $R$. We will use
the notation
\be
\nuc \equiv \delc/S^{1/2} ~~;~~ \nuv \equiv \delv/S^{1/2} ~~;~~
\nuT \equiv \delT/S^{1/2}\,.
\la{nu-def}
\ee
In the following we will frequently refer to the variance $S$ as
``time'', in keeping with standard terminology in excursion set
theory. Statistically we are therefore interested in the probability
distributions of first crossing (f.c.) times, or f.c. \emph{rates}
(conditional or otherwise), which will be denoted by 
\Cal{F}\ with appropriate subscripts. The multiplicity $f$ which
appears in the mass function \eqref{numbercounts} is related to
\Cal{F}\ by $f = 2S\Cal{F}$. SvdW showed that the conditional
f.c. rate of the void barrier, accounting for the void-in-cloud issue
and assuming Gaussian initial conditions, is given by 
\be
\Fsw(S) = \sum_{j=1}^\infty \frac{j\pi}{\delT^2}\sin\left(
j\pi\delv/\delT\right) e^{-j^2\pi^2S/2\delT^2} \,,
\la{FsvdW-inf}
\ee
whose Laplace transform is
\be
\lsw(s) = \int_0^\infty dS e^{-sS}\Fsw(S) = \frac{\sinh\left(
  \sqrt{2s}\delc\right)}{ \sinh\left(\sqrt{2s}\delT\right) } \,.
\la{LsvdW}
\ee
SvdW derived this result based on probabilistic arguments and a clever
use of Laplace transforms. This result is also known in the condensed
matter literature on first crossing problems, and can be found derived
e.g. in \Cite{Redner}, based on a solution of the Fokker-Planck 
equation  
\be
\p_S\Pi = \frac{1}{2}\p^2_\del\Pi\,,
\la{FP-eq}
\ee
where $\Pi(\del,S)$ is the probability density for a diffusing
particle in the presence of absorbing barriers at $\del=\delc$ and
$\del=-\delv$. 

In the following we will be mainly interested in the
large mass or small $S$ limit of \Fsw. As it stands, the expression
\eqref{FsvdW-inf} is not particularly useful in this limit, since for 
small $S$ an increasing number of terms become important\footnote{The 
  expression \eqref{FsvdW-inf} is more useful in the \emph{large} time
  ($S\to\infty$) limit, which is typically encountered in condensed
  matter systems \cite{Redner}.}.  It is very useful therefore to
recast this expression in a form that is better behaved as
$S\to0$. This is not hard to do, and we show in Appendix 
\ref{app-FsvdW-resum} that the conditional f.c. rate \Fsw\ can be
written as  
\be
\Fsw(S) = \frac{1}{(2\pi)^{1/2}S}\sum_{j=-\infty}^\infty(\nuv -
2j\nuT)e^{-(\nuv-2j\nuT)^2/2} \,.
\la{FsvdW}
\ee
(One can check that the series above is also identical to the one
in Eqn. 46 of Lam \etal\ \cite{Lam:2009nd}, with our $j=1$ term
corresponding to their $n=1$, our $j=-1$ to their $n=2$, and so on.)
As $S\to0$, the $j=0$ term rapidly becomes the most important and the
resulting multiplicity reduces to 
\be
S\to0 ~~:~~ \fsw(\nuv,\nuT) = 2S\Fsw(S)\rightarrow
\sqrt{\frac{2}{\pi}} \nuv\ e^{-\nuv^2/2} = f_{\rm PS}(\nuv)\,,
\la{FsvdW-Sto0}
\ee
which is just the 1-barrier Press-Schechter result for the void
barrier. This is not surprising, since for small times it becomes
increasingly unlikely for a trajectory to cross the halo barrier
\emph{and} return to cross the void barrier, reducing the result to a
single barrier one. Figure 7 of \Cite{Sheth:2003py} illustrates this
effect, with the void-in-cloud effects becoming significant only at
$\nuv\lesssim1.5$ or so.

In our non-Gaussian extension we would like to address the question of
whether this intuition continues to remain true, or whether the
void-in-cloud issue is now relevant at very large radii. If we
simply assume that the Gaussian reasoning still holds, then the large
radius end should be describable as a single barrier
problem. This was essentially the reasoning of Kamionkowski 
\etal\ \cite{Kamionkowski:2008sr}, who applied the Press-Schechter
approach to a single void barrier. If we follow this reasoning, then
\emph{at the least} the non-Gaussian void multiplicity $f_{\rm voids}$
should incorporate the single barrier effects discussed in
\Cites{Maggiore:2009rx,D'Amico:2010ta}, and we should expect that 
$f_{\rm voids}$ is given by replacing $\delc\to-\delv$ in the
expression derived by D'Amico \etal\ \cite{D'Amico:2010ta} (their
Eqn. 68 with $\kappa=0$ and $a=1$ to remain within the sharp-k filter
and fixed barrier approximations we are using here). 

In this paper we will address this issue rigorously using path
integral techniques. We will see that, while the void-in-cloud issue
introduces some technical complications, the end result \emph{is} in
fact that at sufficiently large radii one can treat the problem using a
single barrier. For smaller radii, we will see that the effects of the
void-in-cloud issue are not negligible but can be described using a
simple yet accurate approximation. Our results are broadly in
agreement with those of Lam \etal\ \cite{Lam:2009nd}, although our
final expression for the void multiplicity is different from theirs
(see below for a more detailed comparison with their work). We
begin in the next section by introducing some path integral
machinery. 

\section{Deriving the SvdW result from path integrals}
\la{SvdWpathinteg}
\noindent
Our goal in this section is to reproduce using path integrals the SvdW
result which assumes Gaussian initial conditions. This will allow us
to rigorously extend the result to the non-Gaussian
case. Unfortunately the calculations we will end up doing are
technically rather involved. It is instructive therefore to first
go through an intuitive derivation of \Fsw.

Consider the probability distribution $\Pi_{\rm allowed}(\del,S)$ for
the location of the diffusing particle, which solves the Fokker-Planck
equation \eqref{FP-eq} in the presence of two absorbing barriers and
is nonzero in the ``allowed'' region $-\delv<\del<\delc$ between the
barriers. One can interpret the rate \Cal{F}\ at time $S$ as the
amount of probability leaking across the chosen barrier per unit
time. This is similar to constraining the probability to
stay within the barriers until time $S$ and then letting it evolve
freely, as if the barriers were removed at time $S$. The rate at time
$S$ across let's say the void barrier can then be computed as a
derivative of the probability that has leaked across this barrier at
$S$, i.e.  
\be
\Cal{F}_{\rm intuitive}(S) = \p_S\int_{-\infty}^{-\delv} d\del \,
\Pi(\del,S)\,. 
\la{F-intuitive}
\ee
Here $\Pi$ represents the \emph{unconstrained} probability density,
which also satisfies the same Fokker-Planck equation \eqref{FP-eq},
with the ``initial'' condition that at time $S$ we have $\Pi=\Pi_{\rm
  allowed}$. Using the Fokker-Planck equation one can simplify the
expression above to find
\be
\Cal{F}_{\rm intuitive}(S) = \frac{1}{2}\partial_\del \Pi_{\rm
  allowed}(\del,S) \bigg\vert_{\del=-\delv} \,. 
\la{Fsw-simple}
\ee
The solution to the Fokker-Planck equation which vanishes at the
two barriers $\del=\delc$ and $\del=-\delv$ and starts from a Dirac
delta at the origin at the initial time is simply an infinite sum of
Gaussians with shifted mean values,
\begin{equation}
  \Pi_{\rm allowed}(\del,S) =
  \sum_{j=-\infty}^{+\infty}\frac{1}{\sqrt{2\pi S}} 
  \bigg\{\exp\bigg[-\frac{(\del+2j\delT)^2}{2S}\bigg] -
  \exp\bigg[-\frac{(\del-2j\delT+2\delv)^2}{2S}\bigg]\bigg\} \,.
\la{FP-soln-gaussSum}
\end{equation}
This is easy to verify by inspection, since each Gaussian is
separately a solution of the Fokker-Plank equation \eqref{FP-eq}, and
by construction the infinite sum satisfies the boundary conditions
$\Pi(-\delv,S)= \Pi(\delc,S)=0$ (as can be immediately verified for 
$(-\delv)$, while for $\delc$ it is enough to shift $j$ to $j+1$ in the
second Gaussian). Straightforward algebra then shows that the
conditional f.c. rate computed according to \eqn{Fsw-simple} is
precisely the series given in \eqn{FsvdW},
\be
\Cal{F}_{\rm intuitive}(S) = \Fsw(S)\,.
\la{Fsw-equals-Fintuitive}
\ee
These ideas can be made more rigorous using the language of path
integrals. The path integral approach to excursion sets for computing
the halo multiplicity was developed by Maggiore \& Riotto (MR) in a 
series of papers
\cite{Maggiore:2009rv,Maggiore:2009rw,Maggiore:2009rx}, and was   
improved upon by D'Amico \etal\ \cite{D'Amico:2010ta} for the 
non-Gaussian case. In the following we will mainly refer to the
techniques developed in \Cite{Maggiore:2009rv}, restricting ourselves
to the simplest case of a fixed barrier and a sharp-k filter (so that
one is dealing with a Markovian stochastic process). The basic
quantity one deals with is the probability distribution function
$W(\del_0;\{\del_k\}_n;S)$ for a discrete random walk with $n$ steps
$\{\hat\del_k\}|_{k=1}^n$, where $\hat\del_k$ denotes the matter
density contrast smoothed on a scale $R_k$ corresponding to a variance
$S_k$, with steps of equal spacing $\Del S$ in the variance starting
at $S_0$ with corresponding density contrast $\del_0$ (both of which
we will assume to be zero) and with the last step denoted by
$S_n\equiv S$.  We have 
\be
W(\del_0;\{\del_k\}_n;S) \equiv \avg{\del_{\rm D}(\hat\del_1-\del_1)\ldots 
  \del_{\rm D}(\hat\del_n-\del_n)}\,,
\la{W-def}
\ee  
which for the Gaussian, sharp-k filter case reduces to 
\be
W^{\rm gm} =\prod_{k=0}^{n-1}{\Psi_{\DS}(\del_{k+1}-\del_k)} ~;~~~
\Psi_{\DS}(x) = (2\pi\DS)^{-1/2} e^{-x^2/(2\DS)}\,,
\la{Wgm}
\ee
with the superscript standing for ``Gaussian, Markovian''. Throughout
our calculations we will deal with the path integral of this
probability density over the ``allowed'' region -- in the single
barrier case that MR considered this would be the region
$\del_k<\delc$, $1\leq k\leq n-1$, while in our two barrier problem it
will be the region $-\delv<\del_k<\delc$. We will therefore be
interested in objects of the type
\be
\Pi_{\DS}(\del_0,\del_n;S) = \int_{\rm allowed}d\del_1\ldots
d\del_{n-1}\, W(\del_0;\{\del_k\}_n;S) \,.
\la{Pi-prototype}
\ee
This object is the probability density for the diffusing
``particle'' to remain inside its allowed region until time $S$.
Our goal will be to calculate the rate at which this probability leaks
out of the allowed region (i.e. the rate at which the ``particles''
escape) across one of the boundaries. Making these 
ideas rigorous requires us to introduce some technical aspects of the
path integrals, and it will be easiest to do this in the more familiar 
single barrier case. Let us therefore begin with a brief recap of the
(Gaussian) MR calculation for halos in the case of a fixed barrier and
a sharp-k filter. 

\subsection{Recap of the single barrier problem}
\la{recap1bar}
\noindent
As mentioned above, it is useful to define the constrained probability
density $\Pi^{\rm gm}_{1{\rm -bar,}\DS}(\del_0,\del_n;S)$ for the
density contrast at step $n$, given that at all previous steps the
walk remained below the barrier, so that 
\be
\Pi^{\rm gm}_{1{\rm -bar,}\DS}(\del_0,\del_n;S) =
\int_{-\infty}^{\delc}d\del_1\ldots d\del_{n-1} W^{\rm
  gm}(\del_0;\{\del_k\}_n;S) \,. 
\la{Pi-gm-1bar}
\ee
MR showed \cite{Maggiore:2009rv} that the continuum limit of this
quantity recovers the result of Bond \etal\ \cite{Bond:1990iw},
\be
\Pi^{\rm gm}_{1{-bar,}\DS\to0}(\del_0,\del;S) = \Pi_{\rm
  Bond}(\del_0,\del;S) \equiv  \frac{1}{\sqrt{2\pi S}} \left(
e^{-(\del-\del_0)^2/2S} - e^{-(2\delc-\del-\del_0)^2/2S} \right) \,,
\la{Pi-Bond}
\ee
being the solution of the Fokker-Planck equation \eqref{FP-eq}
with initial condition $\Pi(\del_0,\del;0)=\del_{\rm D}(\del-\del_0)$
and boundary conditions $\Pi(\del_0,\delc;S) = 0 =
\Pi(\del_0,\del\to-\infty;S)$. We will go through the derivation of a
similar result for the two barrier case below. 

We are looking for the distribution of the f.c. time  $\hat S_{\rm c}$
at which the random walk first crosses the barrier \delc. This can be 
constructed as follows. Start with the cumulative probability $P(\hat
S_{\rm c}>S)$ that $\hat S_{\rm c} > S$, which is the same as the
probability that the barrier has \emph{not} been crossed until time
$S$, i.e. 
\be
P(\hat S_{\rm c}>S) = \lim_{\DS\to0} \int_{-\infty}^{\delc} d\del_n  
\Pi^{\rm gm}_{1{\rm -bar,}\DS}(\del_0,\del_n;S)\,. 
\la{fc-cumul-MR}
\ee
In the continuum limit this quantity is straightforward to compute
using \eqn{Pi-Bond}. The f.c. rate is then just the negative
derivative of $P(\hat S_{\rm c}>S)$ w.r.t $S$,  leading to the
multiplicity 
\be
f_{\rm PS}(\nuc) = 2S\Cal{F}_{\rm PS}(S) = -2S\p_SP(\hat S_{\rm c}>S)
= \sqrt{\frac{2}{\pi}} \nuc\ e^{-\nuc^2/2} \,, 
\la{rate-1bar}
\ee
which is the celebrated Press-Schechter result (accounting for the
infamous factor of 2), where \nuc\ was defined in \eqn{nu-def}.

On passing to the two barrier problem, in order to
calculate the required constrained f.c. rate, the cumulative
probability we need is $P(\hat S_{\rm v}>S,\hat S_{\rm c}>\hat S_{\rm
  v})$, which is the probability that the void barrier is first crossed
\emph{before} the halo barrier is ever crossed. In this object, the
conditioning of the  stochastic variable $\hat S_{\rm v}$ is thus on
\emph{another} stochastic variable $\hat S_{\rm c}$ which is the first
crossing time for the halo barrier, and this complicates
the issue. In the Gaussian case of course one can solve the problem
the SvdW way, without resorting to path integrals. We are ultimately
interested in the non-Gaussian case though, and we therefore
explore a path integral derivation of the Gaussian SvdW result, which
we can generalize later to the non-Gaussian case. 

\subsection{\Fsw\ from a path integral analysis}
\la{FSvdW-PIdetails}
The basic trick we employ is to exploit the discretized nature of the
path integral, and to pass to the continuum limit carefully. To keep
the discussion as clear as possible, we relegate several technical
derivations to the Appendices. As in the single barrier case, it is
useful to construct the constrained probability density $\Pi^{\rm 
  gm}_{\DS}(\del_0,\del_n;S_n)$ for the density contrast at 
time step $n$, given that the walk has not crossed \emph{either}
barrier at any previous time step,  
\be
\Pi^{\rm gm}_{\DS}(\del_0,\del_n;S_n) =
\int_{-\delv}^{\delc}d\del_1\ldots d\del_{n-1} W^{\rm
  gm}(\del_0;\{\del_k\}_n;S_n)\,. 
\la{Pi-gm-2bar}
\ee
We will need to compute this quantity and its integrals under various 
limits and assumptions, which we will come to presently. To begin
with, note that with a discretized variance parameter $S$ one
can explicitly construct a probability (rather than a density) for the
constrained first crossing of the void barrier to occur at a
\emph{specific} time step $n$. Namely, the integral
$\int_{-\infty}^{-\delv}d\del_n \Pi^{\rm gm}_{\DS}(\del_0,\del_n;S_n)$
is the probability that the walk did not cross either barrier for the first
$n-1$ steps, and crossed the void barrier at step $n$. This is the
same as the probability that the first crossing of the void barrier is
at step $n$, and that the halo barrier has not yet been
crossed. Using this, the cumulative probability $P(\hat S_{\rm
  v}>S,\hat S_{\rm c}>\hat S_{\rm v})$ is obtained by summing over all 
possible choices of the final step $n$ that have $S_n>S$, and then
passing to the continuum limit,
\be
P(\hat S_{\rm v}>S,\hat S_{\rm c}>\hat S_{\rm v}) = \sum_{S_n > S}
\int_{-\infty}^{-\delv}d\del_n \Pi^{\rm gm}_{\DS}(\del_0,\del_n;S_n)
\to \int_S^\infty d\ti{S} \lim_{\DS\to0}\frac{1}{\DS}
\int_{-\infty}^{-\delv}d\del_n \Pi^{\rm
  gm}_{\DS}(\del_0,\del_n;\ti{S})\,. 
\la{cumul-prob}
\ee
The required f.c. rate is simply the negative time derivative of
this object, which we can read off as
\be
\Fws(S) = \lim_{\DS\to0}\frac{1}{\DS} \int_{-\infty}^{-\delv}d\del_n
\Pi^{\rm gm}_{\DS}(\del_0,\del_n;S)\,,
\la{FSvdW-pathinteg}
\ee
which involves the integral of $\Pi^{\rm gm}_{\DS}$ on the ``wrong
side'' of the void barrier, $\del_n<-\delv$ (hence the subscript
WS). One can also see why this is the correct object to compute, since
it corresponds to the amount of probability leaking out of the void
barrier in a time interval \DS, divided by \DS. While the integral
itself will vanish in the continuum limit, we will see that it does so
like $\sim\DS$, leaving a finite limit in \eqn{FSvdW-pathinteg} such
that $\Fws(S)=\Fsw(S)$. 

The above arguments may seem like a convoluted way of arriving at the
result. When we move to non-Gaussian initial conditions for the double
barrier however, we are not left with much
choice in the matter. Nevertheless, it would be reassuring to know that the
``wrong side counting'' described above actually works in some other
situation which is under better control. The single barrier
calculation provides us with such a check. In this case, we can apply
exactly the same arguments as above to obtain the f.c. rate,
\emph{and} we have an alternative derivation of the rate due to
MR. As a check therefore, we should find that these two methods lead
to the same answer. In Appendix \ref{app-wrongsidecounting}, we show
that this is indeed the case for \emph{completely general} initial
conditions (i.e. \emph{without} assuming Gaussianity). In Appendix
\ref{app-comparelam} we also compare our approach with that of Lam 
\etal\ \cite{Lam:2009nd}. Their analysis effectively makes the
assumption that the probability distribution
$W(\del_0;\{\del_k\}_n;S)$ defined in \eqn{W-def} is factorisable,
which is certainly true in the Gaussian case for the sharp-k filter
(see \eqn{Wgm}) but is not valid e.g. in the presence of
non-Gaussianities. We show in Appendix \ref{app-comparelam}
that our expression for the rate (both the Gaussian one of
\eqn{FSvdW-pathinteg} as well as its non-Gaussian generalisation which
we discuss in Section \ref{NGvoids} below) is identical to what they
aim to calculate in Eqn. 41 of \Cite{Lam:2009nd}, but \emph{without}
making this assumption of factorisability. With these reassurances, we
proceed to the main two barrier calculation of this section. Since
the calculation is rather technical in nature, the reader who is
willing to take the result $\Fws=\Fsw$ on faith may skip directly to
the non-Gaussian generalization in Section \ref{NGvoids}.

To calculate the integral in \eqn{FSvdW-pathinteg}, we start by
exploiting the factorisability of $W^{\rm gm}$ to write 
\begin{align}
\Pi^{\rm gm}_{\DS}(\del_0,\del;S+\DS) &=
\int_{\del-\delc}^{\del+\delv}dx\Psi_{\DS}(x) \Pi^{\rm
  gm}_{\DS}(\del_0,\del-x;S)\nonumber\\
&=\sum_{n=0}^\infty\frac{(-1)^n}{n!}\p^n_\del\Pi^{\rm
  gm}_{\DS}(\del_0,\del;S) \int_{\del-\delc}^{\del+\delv}dx
\Psi_{\DS}(x) x^n\nonumber\\
&\equiv \sum_{n=0}^\infty\frac{(-1)^n}{n!}\p^n_\del\Pi^{\rm
  gm}_{\DS}(\del_0,\del;S) \Cal{I}^{\DS}_n(\del)\,,
\la{Pi-gm-factorexpand}
\end{align}
where the first equality follows from the definitions of $\Pi^{\rm
  gm}_{\DS}$, $W^{\rm gm}$ and some relabeling of dummy variables, the
second follows from a Taylor expansion and exchanging the orders of
integration and summation, and the last line defines the functions
$\Cal{I}^{\DS}_n(\del)$, which reduce to
\be
\Cal{I}^{\DS}_n(\del) = \frac{\left( 2\DS\right)^{n/2}}{\pi^{1/2}}
\int_{(\del-\delc)/(2\DS)^{1/2}}^{(\del+\delv)/(2\DS)^{1/2}} dy\,
e^{-y^2} y^n \,.
\la{Inofdel}
\ee
\eqn{Pi-gm-factorexpand} is completely general, and holds for arbitrary
\del. The first thing we can check is that this expression implies
that in the continuum limit, $\Pi^{\rm gm}_{\DS=0}(\del_0,\del;S)$
vanishes on \emph{both} barriers $\del=-\delv$ and $\del=\delc$,
\be
\Pi^{\rm gm}_{\DS=0}(\del_0,-\delv;S) = 0 = \Pi^{\rm
  gm}_{\DS=0}(\del_0,\delc;S)\,. 
\la{bndrycondns}
\ee
This follows by Taylor expanding the l.h.s. of
\eqref{Pi-gm-factorexpand} for small \DS\ and comparing
the lowest order terms on both sides, using
$\Cal{I}^{\DS\to0}_0(\delc)=1/2=\Cal{I}^{\DS\to0}_0(-\delv)$. Here we
set \del\ to \delc\ or $(-\delv)$ \emph{before} taking the continuum
limit. This is a subtle point since these operations do not commute,
which is clear from the structure of $\Cal{I}^{\DS}_n(\del)$. This
brings us to the issue of ``boundary layer'' effects. 

In evaluating the integral in \eqn{FSvdW-pathinteg}, we wish to take
the continuum limit \emph{after} integrating, in contrast to what we
did above for $\Cal{I}^{\DS}_0$. The integral therefore depends not
only on the behaviour of $\Pi^{\rm gm}_{\DS}$ far from the 
void barrier, but also on its detailed behaviour in the ``boundary
layer'' where $|\del+\delv|\sim\sqrt{\DS}$. In our wrong side counting 
approach, there are actually two boundary layers we must worry
about. If we define the quantity $\eta$ by    
\be
\eta \equiv \frac{\del+\delv}{\sqrt{2\DS}}\,,
\la{eta-v}
\ee
then the two boundary layers correspond to regions where
$|\eta|\sim\Cal{O}(1)$, with the one discussed above corresponding to
$\eta<0$. We will also deal with the second layer at $\eta>0$, which
is in fact similar to the one discussed by MR in their single barrier
calculation. In principle there are two more boundary layers, on
either side of the halo barrier as well, but we will only need to deal
with the ones near the void barrier. \fig{boundarylayers} illustrates
the situation. 
\begin{figure}
\centering
\includegraphics[height=0.3\textheight]{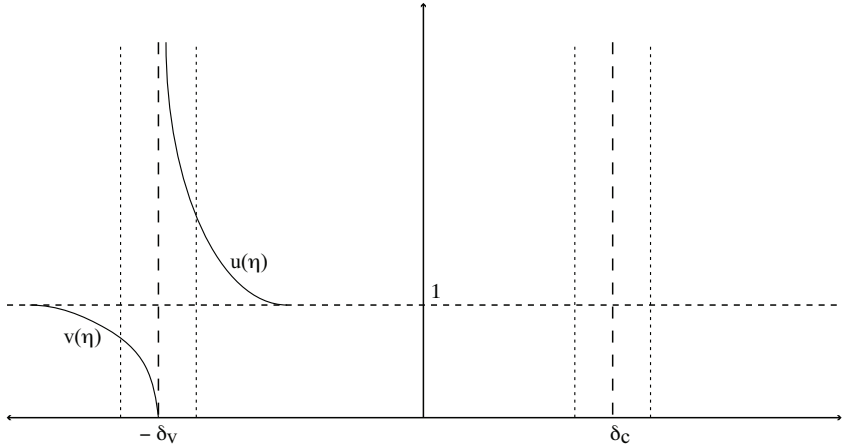}
\caption{\small Schematic view of the two barriers (vertical dashed lines) with
corresponding boundary layers (vertical dotted lines). The boundary
layer functions $v(\eta)$ and $u(\eta)$ are also shown schematically.}
\la{boundarylayers}
\end{figure}
To account for boundary layer effects in the integral in
\eqn{FSvdW-pathinteg}, we use the approach discussed by MR in
\Cite{Maggiore:2009rv}. This involves two steps: first, we calculate
the leading behaviour of $\Pi^{\rm gm}_{\DS}(\del_0,\del;S)$ for small
\DS\ when we hold $\del<-\delv$ to be fixed --
i.e. in the limit of large negative $\eta$. Denote this function by
$C_{\DS}^{(<)}(\del_0,\del;S)$. Next, in order to calculate 
$\Pi^{\rm gm}_{\DS}$ for fixed but small $\DS$ and \emph{arbitrary}
$\del\leq-\delv$ (so that $\eta$ is negative and arbitrary), we
introduce a boundary layer function $v(\eta)$ which we choose to
normalize so that $v(\eta\to-\infty)\to1$, and then write 
\be
\Pi^{\rm gm}_{\DS}(\del_0,\del;S) = v(\eta)
C_{\DS}^{(<)}(\del_0,\del;S) + \ldots\,,
\la{Pi-gm-arbitWS-CDS}
\ee
where the ellipsis indicates terms of higher order in \DS\ than the
leading order in $C_{\DS}^{(<)}$. The function $v(\eta)$ will in
general behave nontrivially in the boundary layer where
$|\eta|\lesssim\Cal{O}(1)$, capturing effects missed by the function 
$C_{\DS}^{(<)}$ on its own. 

Let us now calculate $C_{\DS}^{(<)}$ which can be done by starting with
\eqn{Pi-gm-factorexpand}. In Appendix \ref{app-fsvdw-details-1}, we show
that in this limit of large negative $\eta$ the functions
$\Cal{I}^{\DS}_n(\del)$ reduce to  
\be
\Cal{I}^{\DS}_n(\del<-\delv) = \left(\frac{\DS}{2\pi}\right)^{1/2} 
\left[- e^{-(\del+\delv)^2/2\DS}\left(\del+\delv\right)^{n-1}\left( 1+
  \Cal{O}\left( \frac{\DS}{(\del+\delv)^2} \right) \right) + \ldots
  \right]\,, 
\la{Inofdel-lessdelv}
\ee
where the ellipsis indicates terms which are exponentially
suppressed. Replacing this in \eqn{Pi-gm-factorexpand} we find that
to leading order in \DS, the summation over $n$ can be carried out
exactly using the identity
\be
\sum_{n=0}^\infty \frac{(-1)^n}{n!}\left(\del+\delv \right)^n\p_\del^n
\Pi^{\rm gm}_{\DS}(\del_0,\del;S) = \Pi^{\rm gm}_{\DS}(\del_0,-\delv;S)\,,
\la{Pi-gm-taylor}
\ee
which is not zero for finite \DS. To leading order in \DS, the
l.h.s. of \eqn{Pi-gm-factorexpand} is simply the function
$C_{\DS}^{(<)}(\del_0,\del;S)$ we are looking for, and we can set 
\be
C_{\DS}^{(<)}(\del_0,\del;S) \equiv \left(\frac{\DS}{2\pi}\right)^{1/2}
\frac{e^{-(\del+\delv)^2/2\DS}}{|\del+\delv|} \Pi^{\rm
  gm}_{\DS}(\del_0,-\delv;S)\,,
\la{CDeltaSless}
\ee
where it is understood that we are only interested in the leading
order in \DS. Since the object $\Pi^{\rm gm}_{\DS}(\del_0,-\delv;S)$
is independent of \del\, we will calculate it later. The function
$\Pi^{\rm gm}_{\DS}(\del_0,\del;S)$ for arbitrary $\del\leq-\delv$ can
be written as 
\be
\Pi^{\rm gm}_{\DS}(\del_0,\del\leq-\delv;S) = \frac{1}{2\sqrt{\pi}}
\frac{v(\eta)}{(-\eta)} e^{-\eta^2}\Pi^{\rm gm}_{\DS}(\del_0,-\delv;S)
\,, 
\la{Pi-gm-arbitWS-explicit}
\ee
where we have expressed $C_{\DS}^{(<)}$ in terms of $\eta$. 
In the other limit of fixed \DS\ but $\del\to-\delv$ from below, due
to continuity of $\Pi^{\rm gm}_{\DS}(\del_0,\del;S)$ we have 
\be
v(\eta)\to2\sqrt{\pi}(-\eta) ~~ \textrm{as  } \eta\to0^-\,.
\la{v-etatozero}
\ee
Let us focus now on the integral in \eqn{FSvdW-pathinteg}. It is not
hard to see that this can be re-expressed as an integral over $\eta$
from $-\infty$ to $0$, and reduces to
\be
\int_{-\infty}^{-\delv}d\del \,\Pi^{\rm gm}_{\DS}(\del_0,\del;S) =
\left(\frac{\DS}{2\pi}\right)^{1/2} \left[
  \int_{-\infty}^0\frac{d\eta}{(-\eta)}v(\eta)e^{-\eta^2} \right]
\Pi^{\rm gm}_{\DS}(\del_0,-\delv;S)\,.
\la{Pi-gm-WSinteg}
\ee
The integral in square brackets depends strongly on details of the
function $v(\eta)$, and is in fact dominated by contributions from the
boundary layer where $|\eta|\lesssim1$ due to the exponential
cutoff. We do not know of any way to calculate this object from first
principles. However, this integral is a \emph{finite constant} since
the integrand is well behaved as $\eta\to0^-$ (see \eqn{v-etatozero})
and is exponentially cutoff as $\eta\to-\infty$. Additionally, this
constant only depends on the properties of the \emph{single} void
barrier. We will see below that it can then be fixed by using our
knowledge of the single barrier calculation.

The calculation of \Fws\ will be complete once we evaluate the
object $\Pi^{\rm gm}_{\DS}(\del_0,-\delv;S)$ (which we need only to
leading order in \DS). We start by returning to
\eqn{Pi-gm-factorexpand}, where we are now interested in 
the limit $\del\to-\delv$ from \emph{above} at fixed \DS\ -- i.e as
$\eta\to0^+$. As we shall see next, this will allow us to exploit the
properties of the continuum probability density $\Pi^{\rm gm}_{\DS=0}$
which is non-zero in $-\delv<\del<\delc$ and is moreover
calculable. As one might expect, once again we need to account for a
boundary layer, this time the one where $\eta$ is \emph{positive} and
of order unity. This calculation closely mimics that of a similar
object in the single barrier case \cite{Maggiore:2009rv}. We
essentially repeat the previous analysis of $\Pi^{\rm
  gm}_{\DS}(\del_0,\del;S)$, now on the other side of the void
barrier. Once again we define a function
$C^{(>)}_{\DS}(\del_0,\del;S)$ which reproduces the leading behaviour 
in \DS\ of $\Pi^{\rm gm}_{\DS}(\del_0,\del;S)$ when we hold
$\del\in(-\delv,\delc)$ fixed and let $\DS\to0$. This time we can easily see
that $C^{(>)}_{\DS}(\del_0,\del;S)$ must simply be the continuum limit
probability density 
\be
C^{(>)}_{\DS}(\del_0,\del;S) \equiv \Pi^{\rm
  gm}_{\DS=0}(\del_0,\del;S)\,. 
\la{CDeltaSgreater}
\ee
Introducing a second boundary layer function $u(\eta)$, we can express
$\Pi^{\rm gm}_{\DS}(\del_0,\del;S)$ for arbitrary
$\del\in[-\delv,\delc)$ as
\be
\Pi^{\rm gm}_{\DS}(\del_0,\del;S) = u(\eta)
C^{(>)}_{\DS}(\del_0,\del;S)\,.
\la{Pi-gm-arbit>-delv}
\ee
With $C^{(>)}_{\DS}(\del_0,\del;S)$ defined to be smooth in the
boundary layer where $\eta=\Cal{O}(1)$, we can obtain its leading
behaviour in \DS\ by a simple Taylor expansion, which leads to
\begin{align}
C^{(>)}_{\DS}(\del_0,\del;S) &= \Pi^{\rm
  gm}_{\DS=0}(\del_0,-\delv+\eta\sqrt{2\DS};S) \nonumber\\
&=\eta\sqrt{2\DS} \,\p_\del\Pi^{\rm gm}_{\DS=0}(\del_0,\del=-\delv;S) 
+\ldots\,, 
\la{CDeltaSgreater-expand}
\end{align}
where we used the fact that the continuum limit function $\Pi^{\rm
  gm}_{\DS=0}$ vanishes when evaluated at the barrier. Finally, we can
write the leading order form of $\Pi^{\rm gm}_{\DS}(\del_0,-\delv;S)$
by taking the limit $\eta\to0^+$ in \eqn{Pi-gm-arbit>-delv} to get
\be
\Pi^{\rm gm}_{\DS}(\del_0,-\delv;S) = \gamma\,\sqrt{2\DS}\,
\p_\del\Pi^{\rm gm}_{\DS=0}(\del_0,\del=-\delv;S) ~~;~~ \gamma
\equiv \lim_{\eta\to0^+}\eta\,u(\eta)\,.
\la{Pi-gm-arbit>explicit}
\ee
Here $\gamma$ is another constant which depends on details of the
boundary layer near the void barrier, and is in fact very similar to
the constant which MR compute in \Cite{Maggiore:2009rv} (see their Eqn
78). We will however simply combine it with the other unknown constant
which appears in \eqn{Pi-gm-WSinteg} and fix it below by appealing to
the single barrier problem. Plugging the above relation into
\eqn{Pi-gm-WSinteg}, we see that the integral in \eqn{FSvdW-pathinteg}
is proportional to \DS\ at the leading order. Putting everything
together, the two barrier conditional f.c. rate is then given by  
\be
\Fws = \Cal{A}\,\frac{1}{2} \p_\del\Pi^{\rm
  gm}_{\DS=0}(\del_0,\del=-\delv;S) 
~~;~~ \Cal{A} \equiv \frac{2\gamma}{\sqrt{\pi}}
\,\int_{-\infty}^0\frac{d\eta}{(-\eta)}v(\eta)e^{-\eta^2} \,.  
\la{FSvdW-derived}
\ee
We now need the derivative of the continuum limit solution between the
barriers. If we hold $\del\in(-\delv,\delc)$ fixed and take the limit
$\DS\to0$ then it is easy to see, by expanding both sides of
\eqn{Pi-gm-factorexpand} in powers of \DS\ and equating the lowest
order terms, that the continuum limit function satisfies the
Fokker-Planck equation \eqref{FP-eq} with boundary conditions
\eqref{bndrycondns} and initial condition a Dirac delta centered at
$\del_0$ (which we set to zero for convenience).  In other words, our
rigorous derivation has reproduced \eqn{Fsw-simple}, upto a constant
factor. Since we saw earlier that the expression \eqref{Fsw-simple} is
the same as \Fsw\ in \eqns{FsvdW} and \eqref{FsvdW-inf}, we have
\be
\Fws(S) = \Cal{A}\,\Fsw(S)\,.
\la{Fws-vs-Fsw}
\ee
%
We now show that the constant \Cal{A}\ must be unity. It is clear
from the derivation above that \Cal{A}\ depends only on the properties
of the void barrier. In particular, sending the halo barrier to
infinity, $\delc\to\infty$, would not change the value of \Cal{A}. This,
however, is precisely the limit in which one is solving the
\emph{single} barrier problem. We already know from Appendix
\ref{app-wrongsidecounting} that the single barrier wrong side
counting argument for the f.c. rate is equivalent to the MR
derivation, i.e. $\Cal{F}_{\rm WS, 1-bar} = \Cal{F}_{\rm MR,
  1-bar}$. Additionally, a calculation identical to the one above (but
in the limit of $\delc\to\infty$) shows that $\Cal{F}_{\rm WS, 1-bar} 
= \Cal{A}\,\Cal{F}_{\rm MR, 1-bar}$, and hence we find $\Cal{A}=1$.

\section{Non-Gaussian voids}
\la{NGvoids}
\noindent
With the path integral machinery in place, it is formally
staightforward (although still somewhat involved in practice) to
extend the calculation to the case of non-Gaussian initial
conditions. We simply replace $W^{\rm gm}$ and $\Pi^{\rm gm}_{\DS}$
from the Gaussian calculation with their appropriately generalized
versions. Namely, we have \cite{Maggiore:2009rx}
\begin{align}
W(\del_0;\{\del_k\}_n;S)  &\equiv \avg{\del_{\rm D}(\hat\del_1-\del_1)
  \ldots \del_{\rm D}(\hat\del_n-\del_n)} \nonumber\\
&= \exp\bigg[ -\frac{1}{3!}\sum_{j,k,l=1}^n \avg{\hat\del_j
    \hat\del_k \hat\del_l}_c \p_j\p_k\p_l \nonumber\\
&\ph{\exp -\frac{1}{3!}\sum_{j,k,l=1}^n}
+  \frac{1}{4!} \sum_{j,k,l,m=1}^n \avg{\hat\del_j \hat\del_k
  \hat\del_l \hat\del_m}_c \p_j\p_k\p_l\p_m + \ldots \bigg]  
W^{\rm gm}(\del_0;\{\del_k\}_n;S)\,,
\la{W-generic}
\end{align}
where $\avg{\hat\del_j \hat\del_k \hat\del_l}_c$, $\avg{\hat\del_j
  \hat\del_k \hat\del_l \hat\del_m}_c$, etc. are the connected
moments correlating different length scales (which all vanish in the
Gaussian case, giving back $W^{\rm gm}$). Using this, we also have 
\be
\Pi_{\DS}(\del_0,\del_n;S_n) = \int_{-\delv}^{\delc}d\del_1\ldots
d\del_{n-1} W(\del_0;\{\del_k\}_n;S_n)\,.
\la{Pi-generic}
\ee
One can now apply exactly the same wrong side counting arguments as in
the Gaussian case (see the application to the single barrier case in
Appendix \ref{app-wrongsidecounting}), and find that the required
non-Gaussian f.c. rate is given by 
\be
\Fng(S) = \lim_{\DS\to0}\frac{1}{\DS} \int_{-\infty}^{-\delv}d\del_n
\Pi_{\DS}(\del_0,\del_n;S)\,.
\la{F-NG}
\ee
We will focus on weak primordial non-Gaussianities (NG)
which are expected to be generated in inflationary models. One 
expects that in this case a scale dependent perturbative treatment of
the NG along the lines discussed by D'Amico
\etal\ \cite{D'Amico:2010ta} for the halo multiplicity should work
well for voids as well, with possible complications due to the
void-in-cloud issue. We proceed as in \Cite{D'Amico:2010ta},
beginning by defining the ``equal time'' correlation
functions\footnote{In terms of the reduced cumulants $\Cal{S}_3$, 
  $\Cal{S}_4$, etc. we have $\eps_1=\sig\Cal{S}_3$,
  $\eps_2=\sig^2\Cal{S}_4$, and so on.} (recall $S=\sig_R^2$), 
\be
\eps_{n-2} \equiv \frac{\avg{\hat\del_R^n}_c}{\sig_R^n} ~~;~~
n\geq3\,,
\la{eps-n}
\ee
which are expected to follow the perturbative hierarchy $\eps_n \sim
\ep^n$ for $n\geq1$, where $\ep\ll1$, in generic inflationary models
(see Appendix \ref{app-primNG} for a brief introduction to primordial
NG). Next we Taylor expand the ``unequal time'' correlators appearing
in  \eqn{W-generic} around the final time $S$ \cite{Maggiore:2009rv}
as e.g. 
\be
\avg{\hat\del_j \hat\del_k \hat\del_l}_c = \sum_{p,q,r=0}^\infty
\frac{(-1)^{p+q+r}}{p! q! r!} 
\Cal{G}_3^{(p,q,r)} (S) (S-S_j)^p (S-S_k)^q (S-S_l)^r \, ,
\la{del3expnn}
\ee
where we introduced the (scale dependent) coefficients
\be
\Cal{G}_3^{(p,q,r)}(S) \equiv \left[ \frac{d^p}{dS^p_j}
  \frac{d^q}{d S^q_k} \frac{d^r}{d S^r_l} \avg{\hat\del(S_j)
    \hat\del(S_k) \hat\del(S_l)}_c \right]_{S_j=S_k=S_l=S} \, , 
\la{G3def}
\ee
and similarly for higher point correlations. Such an expansion works
well when considering large scales for which $S$ is small. In this
work we will restrict ourselves to the effects of at most
$\Cal{G}_3^{(1,0,0)}$, ignoring the higher order correlations which
will be suppressed by powers of $S$ and can be tracked as part of the
theoretical error we make in the calculation \cite{D'Amico:2010ta}. We
also introduce a convenient parametrization of $\Cal{G}_3^{(1,0,0)}$
as in \Cite{D'Amico:2010ta} by defining the function $c_1(S)$ via
\be
\Cal{G}_3^{(1,0,0)}(S) = \frac{1}{2}\eps_1(S) c_1(S) S^{1/2}\,,
\la{c1-def}
\ee
which satisfies the identity
\be
c_1(S) = 1+ \frac{2}{3} \frac{d\ln \eps_1}{d\ln S}\,.
\la{c1-eps1dot}
\ee
We will further follow the analysis of \Cite{D'Amico:2010ta} and
simplify the expression for the f.c. rate \eqref{F-NG} by retaining
equal time correlators in the exponential while linearizing the
unequal time contributions, which is a strategy that works well at
least for the single barrier problem. We then find, to the order we
are interested in,
\begin{align}
\Fng(S)  &= \lim_{\DS\to0} \frac{1}{\DS}
\int_{-\infty}^{-\delv}d\del_n \int_{-\delv}^{\delc} d\del_{n-1}\ldots
d\del_1 \exp \left[ -\frac{1}{3!} \avg{\hat\del_n^3}_c
  \sum_{j,k,l} \p_j\p_k\p_l + \frac{1}{4!} \avg{\hat\del_n^4}_c
  \sum_{j,k,l,m} \p_j\p_k\p_l\p_m+ \ldots \right] \nonumber\\
&\ph{\lim_{\DS\to0} \frac{1}{\DS}
\int_{-\infty}^{-\delv}d\del_n \int_{-\delv}^{\delc} d\del_{n-1}}
\left( 1+ \frac{1}{2}\Cal{G}_3^{(1,0,0)}(S)\sum_j(S-S_j)\p_j
\sum_{k,l}\p_k\p_l +\ldots \right) W^{\rm gm}\,.
\la{F-NG-expanded}
\end{align}
In the single barrier case, the analysis at this stage was simplified
by the existence of a very useful identity which states that, for
\emph{any} function $g(\del_1, \ldots,\del_n)$,
\be
\int_{-\infty}^{\delc}{d\del_1\ldots d\del_n\sum_{j=1}^n{\p_j}g} =
\frac{\p}{\p\delc} \int_{-\infty}^{\delc}{d\del_1\ldots d\del_ng}\,.
\la{deriv-exch-1bar}
\ee
This identity is quite easy to prove, in fact it takes only a little
thought to see why it should be true. Remarkably, there is a similar
identity which turns out to be very useful for the \emph{two} barrier
case also, which is not as obvious to write down as
\eqref{deriv-exch-1bar}. As we show in Appendix
\ref{app-fng-details-1}, for any function $g(\del_1, \ldots,\del_n)$
it is also true that
\be
\int_{-\infty}^{-\delv}d\del_n\int_{-\delv}^{\delc}d\del_{n-1}\ldots
d\del_1 \sum_{j=1}^n{\p_j}g = -\left. \frac{\p}{\p\delv}
\right|_{\delT} \int_{-\infty}^{-\delv}
d\del_n\int_{-\delv}^{\delc}d\del_{n-1} \ldots d\del_1 g\,. 
\la{deriv-exch-2bar}
\ee
Notice that the final derivative holds fixed the sum
$\delT=\delv+\delc$ rather than \delc\ alone. For brevity, throughout
the rest of the paper we will omit the explicit reference to this
fact, and simply write $\p_{\delv}$ in place of
$\p/\p\delv|_{\delT}$. 

We can immediately see why this relation is useful. Consider the equal
time exponentiated derivative operator in \eqn{F-NG-expanded}. The
identity \eqref{deriv-exch-2bar} allows us to pull this entire
operator outside the integral, exactly like in the single barrier
case. The remaining terms involving the integrals and the continuum
limit can now be treated individually. We can recognize the first of
these as simply \Fsw, and defining the second as
\be
\Cal{F}^{(\rm 3,NL)} \equiv \lim_{\DS\to0} \frac{1}{\DS}
\int_{-\infty}^{-\delv}d\del_n\int_{-\delv}^{\delc}d\del_{n-1}\ldots
d\del_1  \frac{1}{2} \Cal{G}_3^{(1,0,0)}(S)\sum_j(S-S_j)\p_j
\sum_{k,l}\p_k\p_l \,W^{\rm gm}\,
\la{F3NL}
\ee
(with the notation (3,NL) standing for 3 point, next to leading),
the result at this order is 
\be
\Fng = e^{(1/3!)\eps_1S^{3/2}\p_{\delv}^3 +
  (1/4!)\eps_2S^2\p_{\delv}^4 + \ldots}\left( \Fsw
  +\Cal{F}^{(\rm 3,NL)}+ \ldots \right) \,.
\la{F-NG-interm}
\ee
The term $\Cal{F}^{(\rm 3,NL)}$ is tricky to evaluate, and in Appendix
\ref{app-fng-details-2} we show that it reduces to
\be
\Cal{F}^{(\rm 3,NL)} = -\frac{1}{4\sqrt{2\pi}}
\eps_1c_1S^{1/2} \p_{\delv}^2 \int_0^S
\frac{d\ti{S}}{\sqrt{S-\ti{S}}}\Fsw(\ti{S})\,,
\la{F3NL-eval}
\ee
where we wrote $\Cal{G}_3^{(1,0,0)}$ in terms of $c_1$. Note that
\Fsw\ does depend on both \delv\ and \delT, even though this is hidden
by our compact notation.  From \eqn{FsvdW} we can write
\be
\Fsw(S) = \frac{1}{\sqrt{2\pi}\,S} \sum_{j=-\infty}^\infty
B_j e^{-B_j^2/2} ~~;~~ B_j \equiv \nuv - 2j\nuT \,,
\la{FsvdW-Bj}
\ee
and the integral over $S$ in \eqn{F3NL-eval} can be evaluated exactly,
giving
\be
\int_0^S \frac{d\ti{S}}{\sqrt{S-\ti{S}}}\Fsw(\ti{S}) =
\sum_{j=-\infty}^\infty \frac{1}{S^{1/2}} e^{-B_j^2/2}\,,
\la{F3NL-integral}
\ee
using which we get an expression for $\Cal{F}^{(\rm 3,NL)}$,
\be
\Cal{F}^{(\rm 3,NL)} = -\frac{1}{\sqrt{2\pi}\,S} \frac{1}{4}\eps_1 c_1
\sum_{j=-\infty}^\infty \left(B_j^2-1\right) e^{-B_j^2/2}\,.
\la{F3NL-final}
\ee
Further, noting that in \eqn{F-NG-interm} the combination
$S^{1/2}\p_{\delv}$ can be written as $\p/\p\nuv|_{\nuT,S} \equiv
\p_{\nuv}$, we can write \Fng\ as
\be
\Fng = \frac{1}{\sqrt{2\pi}\,S} e^{(1/3!)\eps_1\p_{\nuv}^3 +
  (1/4!)\eps_2\p_{\nuv}^4+ \ldots} \sum_{j=-\infty}^\infty
e^{-B_j^2/2} \left( B_j - \frac{1}{4}\eps_1c_1 \left(B_j^2-1\right)
+\ldots \right)\,.
\la{F-NG-almostthere}
\ee
Consider the $j=0$ term, for which
$B_0=\nuv$. The exponential derivative can be computed using the
saddle point approximation as discussed in \Cite{D'Amico:2010ta}, and
gives precisely the single barrier f.c. rate computed there for the
fixed barrier excluding filter effects (see their Eqn. 60 with
$\nu\to-\nuv$), 
\begin{align}
\sqrt{\frac2\pi}\,e^{(1/3!)\eps_1\p_{\nuv}^3 +
  (1/4!)\eps_2\p_{\nuv}^4+ \ldots} &\left[ e^{-\nuv^2/2} \left( \nuv -
\frac{1}{4}\eps_1c_1 \left(\nuv^2-1\right) +\ldots \right) \right]
\nonumber\\ 
&= \sqrt{\frac2\pi}\,\nuv e^{-\frac12\nuv^2\left( 1+\eps_1\nuv/3+ \left(\eps_1^2 -
  \eps_2/3 \right)\nuv^2/4 \right)}\left( 1-
\frac{1}{4}\eps_1\nuv(c_1-4) +\ldots \right)\nonumber\\
&\equiv f_{\rm NG,1-bar}(\nuv) \,.
\la{F-NG-jeq0term}
\end{align}
As discussed in detail in \Cite{D'Amico:2010ta}, the ellipsis in the
second line denotes all terms that are parametrically smaller than the
ones written down, which correspond to terms of order
$\Cal{O}(\ep\nuv^{-1},\ep^2\nuv^2,\ep^3\nuv^5)$, where $\ep\ll1$ is
the parameter controlling the NG via $\eps_n\sim\ep^n$. We will assume
that the largest scales we access satisfy $\ep\nuv^3\sim\Cal{O}(1)$. One
can check that for $\fnl\sim100$ this corresponds to Lagrangian scales
$R\sim25h^{-1}$Mpc, or very large voids (see \eqn{com-vs-Lag}). In
this case one finds that the ignored terms listed above are all of the
same order of magnitude. At all smaller scales, the term of the form
$\ep\nuv^{-1}$ gives the largest theoretical error. We will use this
in our arguments below.  

For $j\neq0$, notice first that the $B_j$ are all linear in \nuv, so
that the effect of a derivative $\p_{\nuv}|_{\nuT}$ is identical to
that of a derivative w.r.t $B_j$. We therefore have
\be
\Fng = \frac{1}{\sqrt{2\pi}\,S} \sum_{j=-\infty}^\infty
e^{(1/3!)\eps_1\p_{B_j}^3 + (1/4!)\eps_2\p_{B_j}^4+ \ldots} \left[
e^{-B_j^2/2} \left( B_j - \frac{1}{4}\eps_1c_1 \left(B_j^2-1\right) 
+\ldots \right) \right]\,.
\la{F-NG-jneq0problem}
\ee
At least formally, one might say that the solution is therefore just a
series of single barrier results, leading to a multiplicity $\fng =
2S\Fng$ given by
\begin{align}
\fng(\nuv,\nuT) &= \sqrt{\frac{2}{\pi}} \sum_{j=-\infty}^\infty B_j
e^{-\frac12 B_j^2\left( 1+\eps_1B_j/3+ \left(\eps_1^2 -
  \eps_2/3 \right)B_j^2/4+\ldots\right)} \left( 1-
\frac{1}{4}\eps_1B_j(c_1-4) +\ldots \right)\nonumber\\
&=\sum_{j=-\infty}^\infty f_{\rm NG,1-bar}(B_j)\,,
\la{f-NG-formal}
\end{align}
with $B_j = \nuv - 2j\nuT $ and $f_{\rm NG,1-bar}$ defined in
\eqn{F-NG-jeq0term}. The problem with this expression is that 
for \emph{any} fixed \nuv, \nuT\ and NG parameters $\eps_1$, $\eps_2$,
etc., for large enough $j$ the terms being ignored in the ellipsis
will become comparable to the ones being retained. This is not such an
important issue for the polynomial NG terms, which on their own would
always be suppressed by the Gaussian factor $e^{-\frac12
  B_j^2}$. This is also the reason why analyses such as those of Lam
\etal\ \cite{Lam:2009nd} and Kamionkowski
\etal\ \cite{Kamionkowski:2008sr}, which are based on the Edgeworth
expansion and therefore have multiplicities of the form $e^{-\nu^2/2}$
multiplying a polynomial in $\nu$, are not susceptible to the
void-in-cloud issue for large enough voids. As D'Amico
\etal\ \cite{D'Amico:2010ta} argued however, the Edgeworth series
results break down when the combination $\ep\nu^3$ becomes of order
unity. The D'Amico \etal\ analysis, which we have used here,
instead effectively resums potentially troublesome terms and leads to
the non-trivial series in the exponential in $f_{\rm
  NG,1-bar}(\nu)$, which can give significantly different results for
the halo multiplicity than the Edgeworth-like analysis (see
e.g. Fig. 5 of \Cite{D'Amico:2010ta}). 

In the two barrier case as well, these exponentiated terms are
expected to be important for the $j=0$ piece at large enough
\nuv. However for $j\neq0$ they are problematic, being a series
in the supposedly small parameter $\ep B_j$. Clearly $\left|\ep
B_j\right|$ becomes larger than unity for large enough $j$ (positive
or negative) at \emph{any} fixed \nuv, and the series expansions for
all $j$ values beyond this point break down. This is not surprising,
considering that this expression depends on the saddle point
approximation, which for each $j$ is only valid provided $\left|\ep
B_j\right|<1$ \cite{D'Amico:2010ta}.  A more intuitive way of
understanding this breakdown is to note that 
the result for the f.c. rate is effectively a series of \emph{single
  barrier} f.c. rates with successively larger barrier heights. Since
each single barrier f.c. rate involves the behaviour of a non-Gaussian
conditional p.d.f. ($\Pi_{\rm 1-bar,\DS=0}$) at the barrier, for 
successively larger $j$ we are effectively sampling further and
further extremes of the non-Gaussian tails of the distributions, which
eventually can no longer be described perturbatively. In practice of
course, we don't expect that the resummation of this series would
dominate the Gaussian suppression to the extent of giving order unity
features in the single barrier f.c. rate. 

In fact one can make a stronger statement based on the limit in which
we artificially send $\delc\to\infty$, which \emph{must} recover the
single barrier result for the void barrier. In the expression
\eqref{f-NG-formal}, sending $\delc\to\infty$ for fixed $\delv$ and 
$S$ is the same as sending $\nuT\to\infty$ at fixed \nuv. In this
limit, if the single barrier result is to be recovered, then
\emph{each} term with $j\neq0$ must individually vanish. For a given
$j$, as $\nuT\to\infty$ we have $B_j\to\pm\infty$ with the sign
depending on the sign of $j$. A given term with $j\neq0$ will then
vanish only if the series $1+\eps_1B_j/3+ \left(\eps_1^2 -
  \eps_2/3 \right)B_j^2/4 +\ldots$ which appears in the exponential,
resums as $|B_j|\to\infty$ into a form which is bounded both above and
below by strictly positive numbers. In other words, the exponential 
suppression of the single barrier multiplicity must qualitatively
remain intact. We will make the mild assumption that the lower bound
on the resummed series is not arbitrarily close to zero but is closer
to unity, which excludes pathological features such as e.g. sharp
repeating spikes in the single barrier multiplicity with a slowly
decreasing maximum height. 

With this we can extend the argument to finite \nuT, by noting that in
$f_{\rm NG,1-bar}(B_j)$ sending $\nuT\to\infty$ for fixed $j$ is the
same as sending $|j|\to\infty$ for fixed \nuT. The previous paragraph
immediately implies that the contribution of terms with increasing
$|j|$ is progressively suppressed. Consider first a term in which the
series expansion has broken down, so that $|\ep B_j|=\alpha_0>1$
(while $\ep\nuv$ is still significantly less than unity). The
arguments above suggest that the quantity $f_{\rm NG,1-bar}(B_j)$
resums to the form 
\be
f_{\rm NG,1-bar}(B_j) \sim \alpha_1B_je^{-\frac12\alpha_2B_j^2} =
\frac1\ep\alpha_1 \alpha_0 e^{-\frac1{2\ep^2}\alpha_2\alpha_0^2}\,,
\la{f-NG-1-bar-resum}
\ee
where $\alpha_1$ and $\alpha_2$ are positive numbers with a possible
mild dependence on $|B_j|$. Let us compare this term with the biggest
term ${\rm err}_0$ we do not calculate in $f_{\rm NG,1-bar}(B_0)$,
which is
\be
{\rm err}_0 = \nuv e^{-\frac12\nuv^2\left(1+\eps_1\nuv/3+
  \left(\eps_1^2 - \eps_2/3 \right)\nuv^2/4 \right)}
\times\Cal{O}\left(\ep\nuv^{-1}\right)  \,.
\la{err0}
\ee
The ratio of these terms is
\be
r \equiv \frac{f_{\rm NG,1-bar}(B_j)}{{\rm err}_0} \sim
\frac{\alpha_1\alpha_0}{\ep^2} e^{-\frac1{2\ep^2} \left(
  \alpha_2\alpha_0^2 - (\ep\nuv)^2\left(
  1+\Cal{O}(\ep\nuv)\right)\right)} \,, 
\la{error-ratio}
\ee
and as long as this ratio is less than unity the error we make by
ignoring $f_{\rm NG,1-bar}(B_j)$ is smaller than the one made by
truncating the series in $f_{\rm NG,1-bar}(B_0)$. We argued above that
the quantity $\alpha_2$ is strictly positive. Moreover we expect
$\alpha_1$ to be of the same order as $\alpha_2$, since both are
proxies for resummed series which are structurally similar.
Since we cannot compute these objects, let us separately analyse the
situations where $\alpha_2$ and $\alpha_1$ are of order unity or are
much smaller. 
\begin{itemize}
\item If $\alpha_2,\alpha_1$ are of order unity, then since
  $\alpha_0>1$ the first term in the exponential is much larger than
  the term containing $(\ep\nuv)^2$, and clearly we have $r\ll1$. 
\item Even if $\alpha_2,\alpha_1$ are not close to unity, in practice $r$
  remains small provided only that $\alpha_2$ is not very close to
  zero, this assumption following from our comment above on the lower
  bound on the exponentiated series. For example if we assume
  $\alpha_2,|\alpha_1|>(\ep\nuv)^2$ then we find  
  \be 
  r<\alpha_0\nuv^2 e^{-\frac12\nuv^2(\alpha_0^2-1 )}\,. 
  \la{r-bound}
  \ee
  Now even if $\alpha_0$ is not much larger than unity the ratio $r$
  will remain small for all interesting values of \nuv. Say
  $\alpha_0\simeq1.5$, then $r$ is already less than unity at
  $\nuv=1$, which is the lower limit for our formalism in any case.
\end{itemize}
To summarize, $j$-values for which the series expansion in $|\ep B_j|$
has formally broken down (even mildly) are not expected to give
contributions larger than the terms which are already being ignored in
the $j=0$ term. 

\vskip 0.05in

Finally, for terms in \eqn{f-NG-formal} with non-zero $j$ where the
series expansion has \emph{not} broken down, one might still expect
that the contribution of these terms is always smaller than ${\rm
  err}_0$. We will now see that this is not quite true, although it is
possible to ignore the $j\neq0$ terms if we only probe the largest
voids. We start by comparing the leading order piece of $f_{\rm
  NG,1-bar}(B_j)$  which is given by $B_je^{-\frac12 B_j^2\left(
  1+\Cal{O}(\ep B_j)\right)}$, with ${\rm err}_0$. The ratio of these
terms is 
\be
\left| \frac{B_je^{-\frac12 B_j^2 \left( 1+\Cal{O}(\ep
    B_j)\right) }}{{\rm err}_0} \right| =
\frac{B_j}{\ep} e^{-\frac{1}{2} \left( B_j^2 - \nuv^2 +\ldots \right)}\,,
\la{err-ratio-2}
\ee
with the ellipsis denoting parametrically smaller quantities. In order
to be able to ignore this term, we need to place an upper bound on the
above ratio and show that this bound is less than unity. Since
this entire discussion is valid only if $\fnl\neq0$ (else we simply
use \fsw), suppose now that we have a lower bound for $|\fnl|$, say
$|\fnl|>1$ in the local model, which translates to $\ep>2\cdot10^{-4}$
or $\ep^{-1}<5000$. Writing $B_j=\nuv\left( 1-
2j\delT/\delv\right)$ with $\delT/\delv\approx5/3$ for
$\delc\simeq1.7$ and $\delv\simeq2.7$,  we can see that $|B_j|>2\nuv$
for all $j\neq0$. Together with the bound on \ep, this allows us to
calculate a minimum value ${\nuv}_{\rm min}$ such that the ratio
\eqref{err-ratio-2} is always less than unity. We do this by setting
$|B_j|=2\nuv$ since all larger values will give smaller ratios, and
find that the ratio \eqref{err-ratio-2} is guaranteed to be less than
unity if $\nuv>{\nuv}_{\rm min}\simeq2.5$ corresponding to voids with 
Lagrangian radius larger than $\sim6.5h^{-1}$Mpc (or comoving radius
larger than $\sim11h^{-1}$Mpc) in a \emph{WMAP}-compatible
$\Lambda$CDM cosmology \cite{Komatsu:2010fb}. If we assume a smaller
lower bound on $|\fnl|$, the value of ${\nuv}_{\rm min}$ will increase
slowly. 

We do not need to be this conservative however. We can also consider
values $\nuv<{\nuv}_{\rm min}$ by retaining an appropriate number of
terms with $j\neq0$, depending on the chosen value of \fnl\ or \ep. In
fact as we will see presently, in practice there is an even simpler
way of accounting for the void-in-cloud effect. Firstly, our arguments
above indicate that at a given $\nuv>1$ we must keep the leading order 
behaviour (schematically $\sim B_je^{-\frac12B_j^2(1+\ep
  B_j+\ldots)}$) of those $j\neq0$ terms for which $|\ep B_j|<1$ and
the ratio in \eqn{err-ratio-2} is larger than unity. (An analysis
similar to the one above shows that the subleading terms can always be
ignored in this case, so that the dominant error is still given by
${\rm err}_0$.) This is straightforward to implement numerically, and
in \fig{fig-ratio} we show the ratio
$\fng(\nuv,\nuT)/\fsw(\nuv,\nuT)$ (solid red) as a function of
$R_{\rm com}$ for $\fnl=\pm100$ at $z=0$ with $\delv=2.72$ and
$\delc=1.686$. 
\begin{figure}[t]
\centering
\subfloat[]{\includegraphics[width=0.475\textwidth]{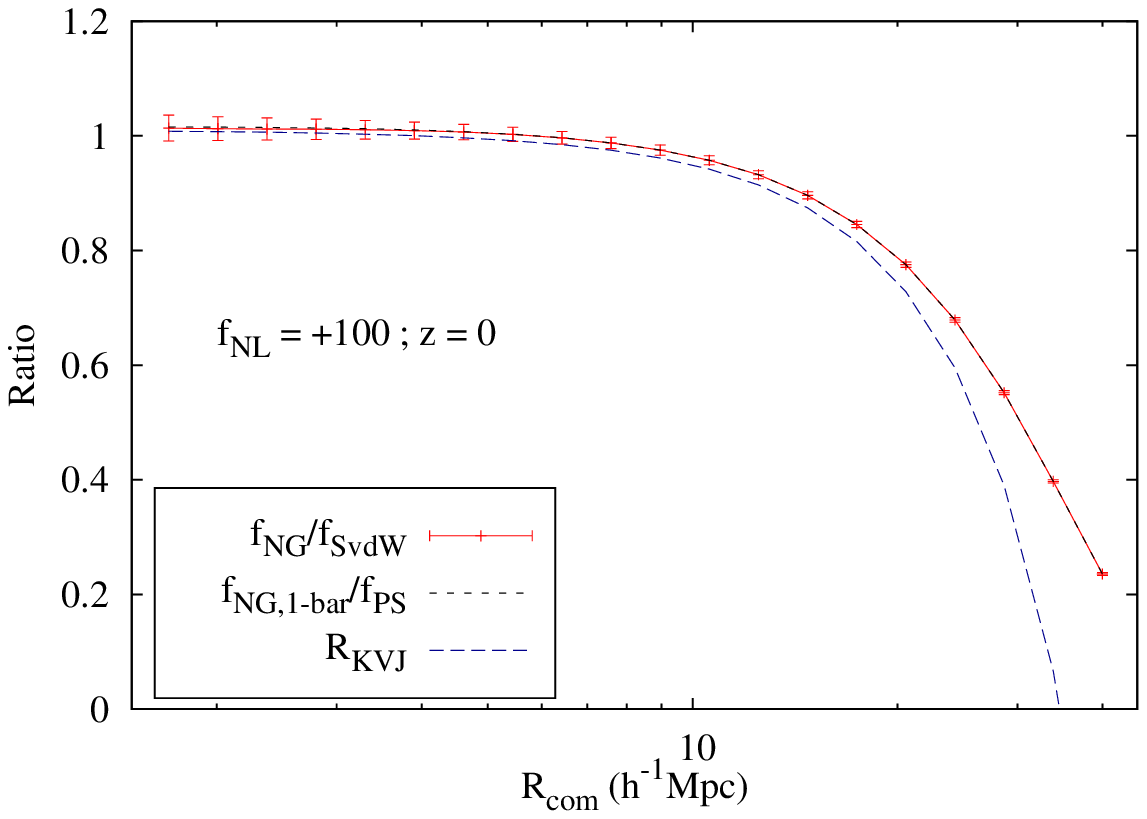}}
\la{fig-ratio-a}
\hspace{0.03\textwidth}
\subfloat[]{\includegraphics[width=0.475\textwidth]{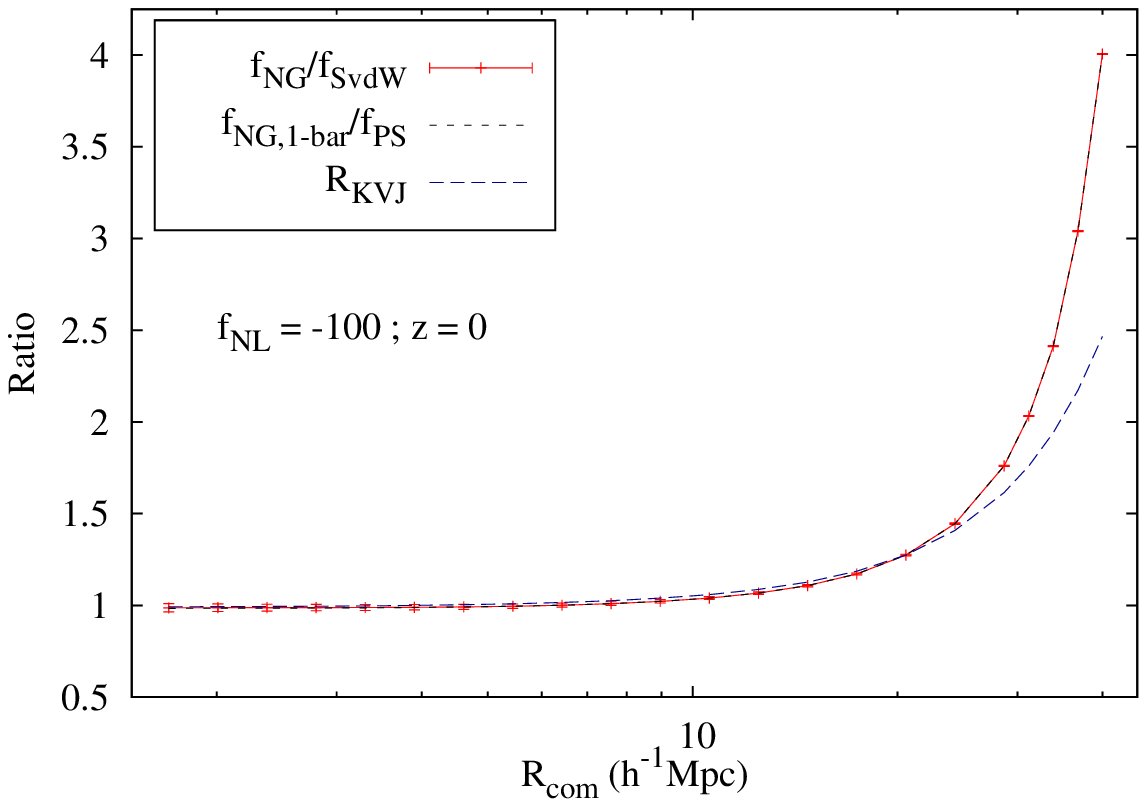}}
\la{fig-ratio-b}
\caption{\footnotesize The ratio of non-Gaussian to Gaussian
  multiplicity for $\fnl=+100$ (left panel) and $\fnl=-100$ (right
  panel) in the local model, at redshift $z=0$. The two barrier and
  single barrier ratios are indistinguishable, as discussed in the
  text. There is a clear departure at large radii from the ratio
  $R_{\rm KVJ}$ proposed by Kamionkowski
  \etal\ \cite{Kamionkowski:2008sr}, which is the one calculated by
  LoVerde \etal\ \cite{LoVerde:2007ri}. The error bars represent
  theoretical errors of order $\Cal{O}(\ep\nuv^{-1})$ as discussed in
  the text.} 
\la{fig-ratio}
\end{figure}
We also plot the ratio of the \emph{single barrier}
functions $f_{\rm NG,1-bar}(\nuv)/f_{\rm PS}(\nuv)$ (short-dashed
black), see \eqn{F-NG-jeq0term}. As we see, these curves are
indistinguishable, and one can also check that even at the smallest
radii we consider, the difference between the curves is less than
$0.5\%$. In fact this result is also easy to see analytically, by
recognizing that the $j\neq0$ terms give increasingly smaller
contributions to the sum in $\fng(\nuv,\nuT)$. One then has
\begin{align}
\frac{\fng(\nuv,\nuT)}{\fsw(\nuv,\nuT)} &= \frac{f_{\rm
    NG,1-bar}(\nuv)+\sum_{j\neq0}f_{\rm NG,1-bar}(B_j)}{f_{\rm
    PS}(\nuv) + \sum_{j\neq0}f_{\rm PS}(B_j)} \nonumber\\
&\approx \frac{f_{\rm NG,1-bar}(\nuv)}{f_{\rm PS}(\nuv)} \left[ 1 +
  \sum_{j\neq0} \left( \frac{f_{\rm NG,1-bar}(B_j)}{f_{\rm
      NG,1-bar}(\nuv)} - \frac{f_{\rm PS}(B_j)}{f_{\rm
      PS}(\nuv)}\right)\right] \,,
\la{ratio-relation}
\end{align}
where we linearized in the $j\neq0$ terms. Now, for large \nuv\ the 
summation in the second line of \eqref{ratio-relation} will be
suppressed simply because of the Gaussian factor
$e^{-\frac12(B_j^2-\nuv^2)}$ in each term. For $\nuv\to1$
this supression will not be very strong at least for small values of
$|j|\neq0$. However, for such terms the \emph{single barrier} ratio
$f_{\rm NG,1-bar}/f_{\rm PS}$ approaches a constant, leading to a
cancellation of the terms in the summation above. In practice
therefore, to extremely good accuracy the two barrier non-Gaussian
multiplicity can be simply written as the product of the single
barrier non-Gaussian ratio with the two barrier \emph{Gaussian}
multiplicity \fsw, 
\be
\fng(\nuv,\nuT) = \sum_{j=-\infty}^\infty f_{\rm NG,1-bar}(B_j)
\approx \left( \frac{f_{\rm NG,1-bar}(\nuv)}{f_{\rm PS}(\nuv)}\right)
\times \fsw(\nuv,\nuT)\,,
\la{finalresult}
\ee
where $f_{\rm NG,1-bar}(\nu)$ was defined in \eqn{F-NG-jeq0term},
$f_{\rm PS}(\nu)$ in \eqn{rate-1bar} and $B_j = \nuv - 2j\nuT $.
Of course this discussion is subject to the caveat that there is
always a theoretical error at least due to the terms we do not compute
in the single barrier multiplicity. For comparison, in 
\fig{fig-ratio} we also show (dashed blue) the ratio proposed by
Kamionkowski \etal\ \cite{Kamionkowski:2008sr}, which (as expected)
deviates from our prediction at large radii. The same will be true of
the ratio calculated by Lam \etal\ \cite{Lam:2009nd} at large
radii. \fig{fig-ratio} also illustrates the complementary nature of the void
abundances as a probe of NG, since the abundance at large radii is
reduced compared to the Gaussian case for positive values of \fnl\ and
vice versa for negative \fnl, which is the opposite of what happens
for halo abundances. 

\begin{figure}[t]
\centering
\includegraphics[width=0.65\textwidth]{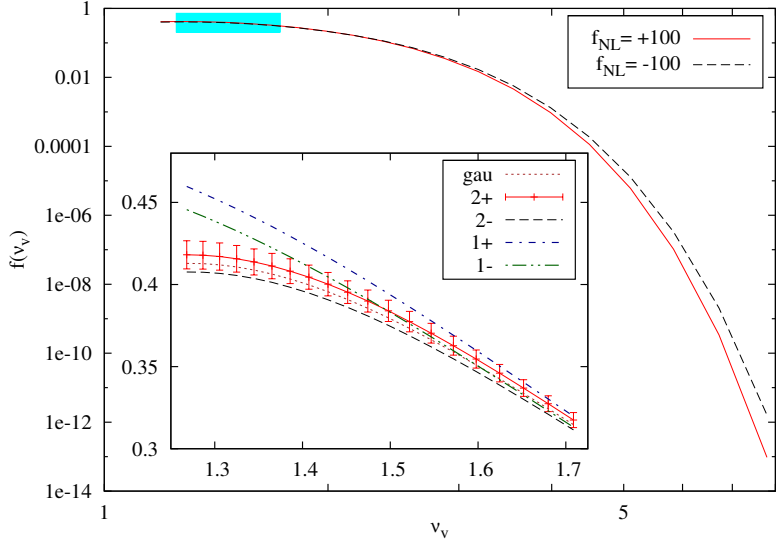}
\caption{\footnotesize The two barrier non-Gaussian multiplicity
  \fng\ as a function of \nuv, for $\fnl=\pm100$. The inset shows a
  zoomed in view of the range highlighted by the cyan (shaded)
  area, corresponding to the range $R_{\rm com}=2$-$4h^{-1}$Mpc. The
  curves in the inset correspond to \fsw\ (gau), $\fng^{(\fnl=\pm100)}$
  ($2\pm$) and $f_{\rm NG,1-bar}^{(\fnl=\pm100)}$ ($1\pm$). The single
  barrier result overestimates the void abundance by order $\sim10\%$
  for \emph{both} signs of \fnl. The correct, two barrier result on
  the other hand displays a behaviour opposite to that at large
  $R_{\rm com}$ : positive \fnl\ slightly enhances the abundance at
  small $R_{\rm com}$. Also shown are the theoretical errors on the
  two barrier result as discussed in the text. For clarity we only show
  these for $\fnl=+100$.}
\la{fig-mult}
\end{figure}
In \fig{fig-mult} we show the non-Gaussian multiplicity as a
function of \nuv, together with a zoomed in view at small comoving
radii ($\nuv\gtrsim1$). We see that ignoring the $j\neq0$
void-in-cloud terms entirely will overpredict the void abundance at
small radii for either sign of \fnl. On the other hand, accounting for
void-in-cloud effects gives a result which is approximately the same
as the \emph{Gaussian} one, with a slight enhancement for positive
\fnl\ and a slight reduction for negative \fnl. Notice that this is
the opposite of what happens at \emph{large} radii. Unfortunately the
magnitude of  the reversed effect at small radii appears to be too
small to be observationally relevant. Finally, in \fig{fig-numdens} we
show the differential comoving number density $dn_{\rm com}/d\log
R_{\rm com}$ defined in \eqn{numbercounts} as a function of comoving
radius $R_{\rm com}$.
\begin{figure}[t]
\centering
\includegraphics[width=0.55\textwidth]{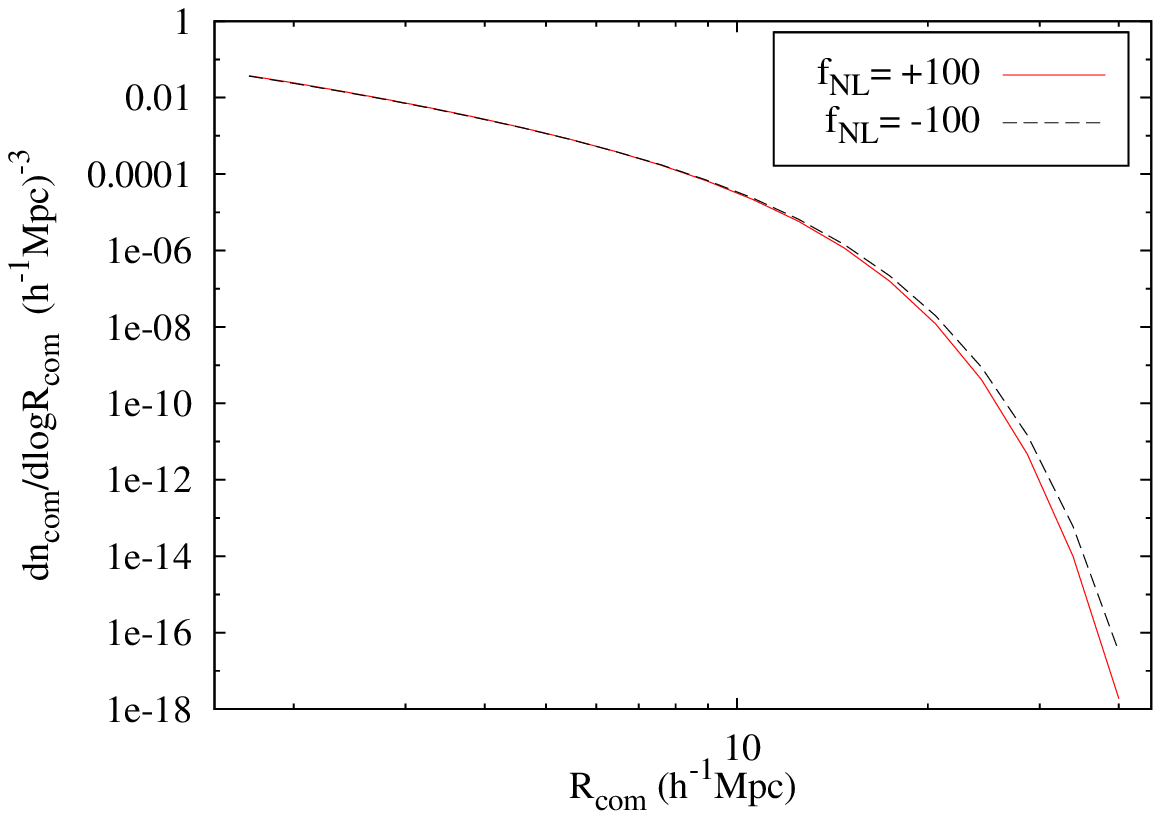}
\caption{\footnotesize Comoving number density defined in
  \eqn{numbercounts}, as a function of comoving radius $R_{\rm com}$
  at redshift $z=0$ for $\fnl=\pm100$.}
\la{fig-numdens}
\end{figure}

\section{Discussion}
\la{discuss}
\noindent
Primordial non-Gaussianity (NG) can be probed by the imprints it
leaves on the late time large scale structure of the universe
by modifying the distribution of matter, which is
manifested for example in the abundance of collapsed objects. In this
paper we have explored a second manifestation of this effect, which is
the abundance of \emph{voids} or underdense regions. While such
calculations have been performed earlier in the literature 
\cite{Kamionkowski:2008sr,Lam:2009nd}, they have been subject at least 
to the caveat that their treatment of the NG was based on a
linearization in \fnl, which is known to potentially misestimate the
abundance of very massive objects \cite{D'Amico:2010ta}. Our
calculation was based on path integral techniques introduced by
Maggiore \& Riotto \cite{Maggiore:2009rv,Maggiore:2009rx}, and
importantly also on the improvements to these techniques developed in 
\Cite{D'Amico:2010ta}. The latter allow us to access a larger range of
length scales than treatments based on linearizing in \fnl, while the
use of path integrals also allows us to carefully account for the 
``void-in-cloud'' issue first pointed out by Sheth \& van de Weygaert
\cite{Sheth:2003py}, which is unique to the case of voids. 
We showed that in the final analysis, the complication introduced by
the void-in-cloud issue can actually be accounted for in a fairly
simple manner, with the void multipicity $\fng(\nuv,\nuT)$ being given
by the approximation in \eqn{finalresult}. Our treatment of the
multi-scale correlations arising from primordial NG is more rigorous
than that of Lam \etal\ \cite{Lam:2009nd} (although these effects are
small), and additionally our explicit expression for the
multiplicity is different from theirs due to the presence of terms
involving the NG functions $\eps_1$, $\eps_2$, etc. in an exponential
(see e.g. \eqn{F-NG-jeq0term}). Our prediction for the void
multiplicity (or number density) is also simple to implement
numerically, especially when combined with the fits for the NG
functions given in \Cite{D'Amico:2010ta}.

It is worth spending a moment to compare the relative merits of using
voids as opposed to halos as a probe of primordial NG. In terms of
absolute numbers, a given survey (or simulation box) will always
contain fewer voids of some Lagrangian scale $R$ than halos of the
corresponding mass scale $M\propto R^3$. This is simply because the
void barrier height \delv\ is larger than the halo barrier height
\delc, leading to a stronger cutoff $\sim e^{-\frac12\nuv^2}$ for the
void abundance. Furthermore on comparing the typical masses of the
largest clusters observed ($M\sim10^{15}h^{-1}\Msol$) with the typical
comoving sizes of the largest voids ($R_{\rm com}\sim 50h^{-1}$Mpc),
one finds that the largest \emph{Lagrangian} length scales probed by
both halos and voids are roughly the same ($R\sim25h^{-1}$Mpc). In
this sense voids would be a poorer statistical probe of primordial NG
than halos. The strength of voids however comes from the
complementarity which is demonstrated in \fig{fig-ratio}, and 
was also highlighted in \Cites{Kamionkowski:2008sr,Lam:2009nd}. In
contrast to the halo abundance which is e.g. reduced compared to the
Gaussian for negative \fnl, the void abundance is enhanced at large
radii, and vice versa for positive \fnl. It will be interesting to see
how these characteristics ultimately play out in determining the
constraining power of voids. 

Our work can be extended in more than one direction. Firstly, our
arguments regarding the terms with large $|j|$ in \eqn{finalresult}
were somewhat qualitative, although we expect them to be robust. In a
future work, we will test these arguments by numerically generating
the appropriate random walks and explicitly determining the
multiplicity using the resulting distribution of conditional first
crossing times \cite{D'Amico:future}. Of course it will also be
interesting to compare our predictions (and those of others) with
full-fledged $N$-body simulations. On the analytical side, it will be
interesting to study the regime $\nuv\gtrsim1$ in more detail, since
apart from void-in-cloud effects it is also likely that triaxial
effects would become important here \cite{Sheth:1999su,Sheth:2001dp}. 
It would be interesting to try and account for such effects (even
approximately) within our framework, perhaps along the lines explored
in \Cite{DeSimone:2010mu} (see also
\Cites{Lam:2009nb,Lam:2009nd}). Finally, an equally interesting avenue
would be to convert our predictions which hold for the dark matter
distribution, to predictions for voids in the \emph{galaxy}
distribution, perhaps by generalizing the treatment of
\Cite{Furlanetto:2005cc} to the non-Gaussian case.

\section*{Acknowledgements}
It is a pleasure to thank Paolo Creminelli, Michele Maggiore, Jorge
Moreno and Licia Verde for useful discussions, and especially Ravi
Sheth for detailed discussions and comments on an earlier draft.

\section*{Appendix}
\appendix
\numberwithin{equation}{section}
\section{\normalsize Primordial non-Gaussianity}
\la{app-primNG}
\noindent
The physics of inflation governs the statistics of the initial seeds of
inhomogeneities which grow into the large scale structure we see
today. These initial conditions can be characterised by the primordial
comoving curvature perturbation $\R(\vec{x})$ (with Fourier transform
$\R(\k)$) which remains constant on superhorizon scales. The function
$\avg{\R(\k_1) \R(\k_2) \R(\k_3)}_c$ (with the subscript denoting the
connected part) is then an example of a function which probes the
physics of the inflationary epoch. By translational invariance, it is
proportional to a momentum-conserving delta function: 
\be
\avg{\R(\k_1) \R(\k_2) \R(\k_3)}_c = (2 \pi)^3
\del_D(\k_1+\k_2+\k_3) 
B_\R(k_1,k_2,k_3) \, ,
\ee
where the (reduced) bispectrum $B_\R(k_1,k_2,k_3)$ depends only on the
magnitude of the $k$'s by rotational invariance.
According to the particular model of inflation, the
bispectrum will be peaked about a particular shape of the triangle. 
The two most common cases are the squeezed (or local) NG,
peaked on squeezed triangles $k_1 \ll k_2 \simeq k_3$, and the
equilateral NG, peaked on equilateral triangles $k_1
\simeq k_2 \simeq k_3$. Indeed, one can define a scalar product of
bispectra, which describes how sensitive one is to a
NG of a given type if the analysis is performed using
some template form for the bispectrum.
As expected, the local and equilateral shapes are approximately orthogonal
with respect to this scalar product~\cite{Babich:2004gb}.
We will now describe these two models in more detail. 
\vskip 0.1in
\noindent
\underline{\it\large The local model:}
\vskip 0.075in
\noindent
The local bispectrum is produced when the NG is generated 
outside the horizon, for instance in the curvaton
model~\cite{Lyth:2002my, Bartolo:2003jx} or in the inhomogeneous
reheating scenario~\cite{Dvali:2003em}. 
In these models, the curvature perturbation can be written in the
following form,
\be
\R(\x) = \R_g(\x) + \frac{3}{5} \fnl^{\rm loc} \left(\R_g^2(\x) -
\avg{\R_g^2}\right) 
+ \frac{9}{25} \gnl \R_g^3(\x) \, , 
\ee
where $\R_g$ is the linear, Gaussian field. We have included also a
cubic term, which will generate the trispectrum at leading order. 
The bispectrum is given by 
\be
B_\R(k_1,k_2,k_3) = \frac{6}{5} \fnl^{\rm loc} \left[
  P_\R(k_1) P_\R(k_2) + {\rm cycl.} \right] \, ,
\ee
where ``cycl.'' denotes the 2 cyclic permutations of the wavenumbers,
and $P_\R(k)$ is the power spectrum given by $P_\R(k)=Ak^{n_s-4}$.
The trispectrum is given by
\begin{multline}
\avg{\R(\k_1) \R(\k_2) \R(\k_3) \R(\k_4)}_c =
(2 \pi)^3 \del_D(\k_1+\k_2+\k_3+\k_4) \\
\times \left[ \frac{36}{25} \fnl^2 \sum_{\substack{b < c \cr a \neq b,
      c}} P_\R(|\k_a + \k_b|) P_\R(k_b) P_\R(k_c) 
+ \frac{54}{25} \gnl \sum_{a<b<c} P_\R(k_a) P_\R(k_b) P_\R(k_c)
\right] \, . 
\end{multline}
%

\vskip 0.1in
\noindent
\underline{\it\large The equilateral model:}
\vskip 0.075in
\noindent
Models with derivative interactions of the inflaton
field~\cite{Alishahiha:2004eh, ArkaniHamed:2003uz, Creminelli:2003iq}
give a bispectrum which is peaked around equilateral configurations,
whose specific functional form is model dependent.
Moreover, the form of the bispectrum is usually not convenient to use
in numerical analyses.
This is why, when dealing with equilateral NG,
it is convenient to use the following parametrization, given
in~\Cite{Creminelli:2005hu},
\begin{equation}
B_\R(k_1,k_2,k_3) = \frac{18}{5} \fnl^{\rm equil} A^2
\Big[ \frac{1}{2 k_1^{4-n_s}  k_2^{4-n_s}}
+ \frac{1}{3 (k_1 k_2 k_3)^{2 (4-n_s)/3}} \\
- \frac{1}{(k_1 k_2^2 k_3^3)^{(4-n_s)/3}} + \text{5 perms.} \Big] \, .
\la{equilateral}
\end{equation}
This is peaked on equilateral configurations, and its scalar product
with the bispectra produced by the realistic models cited above is
very close to one.  Therefore, being a sum of factorizable functions,
it is the standard template used in data analyses. 

To connect the statistics of \R\ with large scale structure, we use
the fact that the excursion set ansatz only requires us to know the
linearly extrapolated present day behaviour of the density contrast
$\del_R$ smoothed on various length scales. We can relate this
quantity to the initial conditions \R\ via the following relations. We
start from the Bardeen potential $\Phi$ on subhorizon scales, given by  
\be
\Phi(\k, z) = - \frac{3}{5} T(k) \frac{D(z)}{a} \R(k) \, ,
\la{Phi}
\ee
where $T(k)$ is the transfer function of perturbations, normalized to
unity as $k \to 0$, which describes the suppression of power for modes
that entered the horizon before the matter-radiation equality (see
e.g. \Cite{Bardeen:1985tr}); and $D(z)$ is the linear growth factor of
density fluctuations, normalized such that $D(z)=(1+z)^{-1}$ in the
matter dominated era. Then, the density contrast field is related to
the potential by the cosmological Poisson equation, which in Fourier
space reads  
\begin{align}
\del(\k, z) &= - \frac{2 a k^2}{3 \Om_{m} H_0^2} \Phi(\k,z)
= \frac{2 k^2}{5 \Om_{m} H_0^2} T(k) D(z) \R(k) \nonumber \\
&\equiv \Cal{M}(k,z) \R(k) \, ,
\la{transfer}
\end{align}
with $\Om_{m}$ the present time fractional density of matter (cold
dark matter and baryons), and $H_0=100h\, {\rm km\,s}^{-1}{\rm 
  Mpc}^{-1}$ the present time Hubble constant. The redshift
dependence is trivially accounted for by the linear growth factor
$D(z)$. Introducing a filter function $W_R(|\x|)$,  the
smoothed density field (around one point, which we take as the
origin) is given by
\be
\del_R(z) = \int \frac{d^3 k}{(2 \pi)^3} \tW(k R) \del(\k,z) \, , 
\la{delR}
\ee
where $\tW(k R)$ is the Fourier transform of the filter function. 
The results of this paper are strictly valid only for a sharp filter
in $k$-space, although a physically more relevant filter would be the
spherical top-hat filter in real space, whose Fourier transform
$\tW(kR)$ is given by 
\be
\tW(y) = \frac{3}{y^3}\left(\sin y - y\cos y \right)\,.
\la{filter-sharpx}
\ee
\begin{figure}[t]
\centering
\subfloat[]{\includegraphics[width=0.47\textwidth]{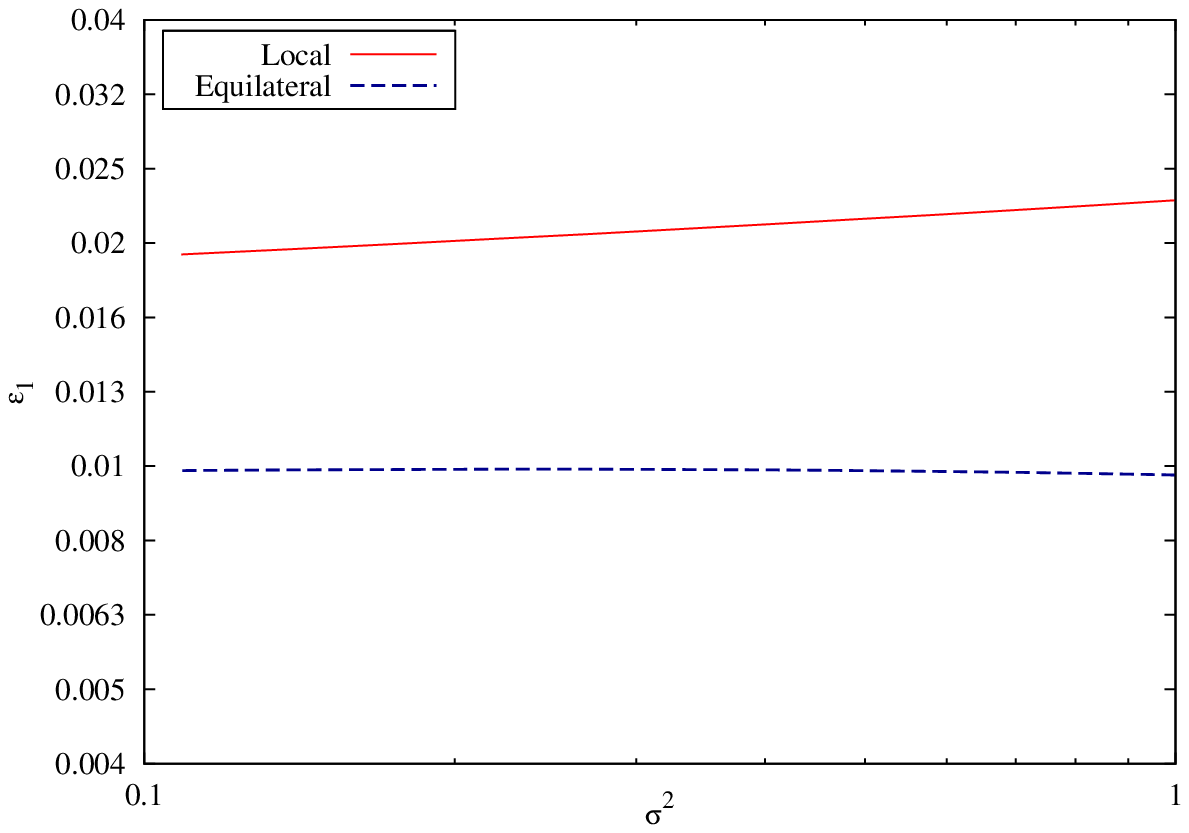}}
\la{fig-eps-a}
\hspace{0.03\textwidth}
\subfloat[]{\includegraphics[width=0.47\textwidth]{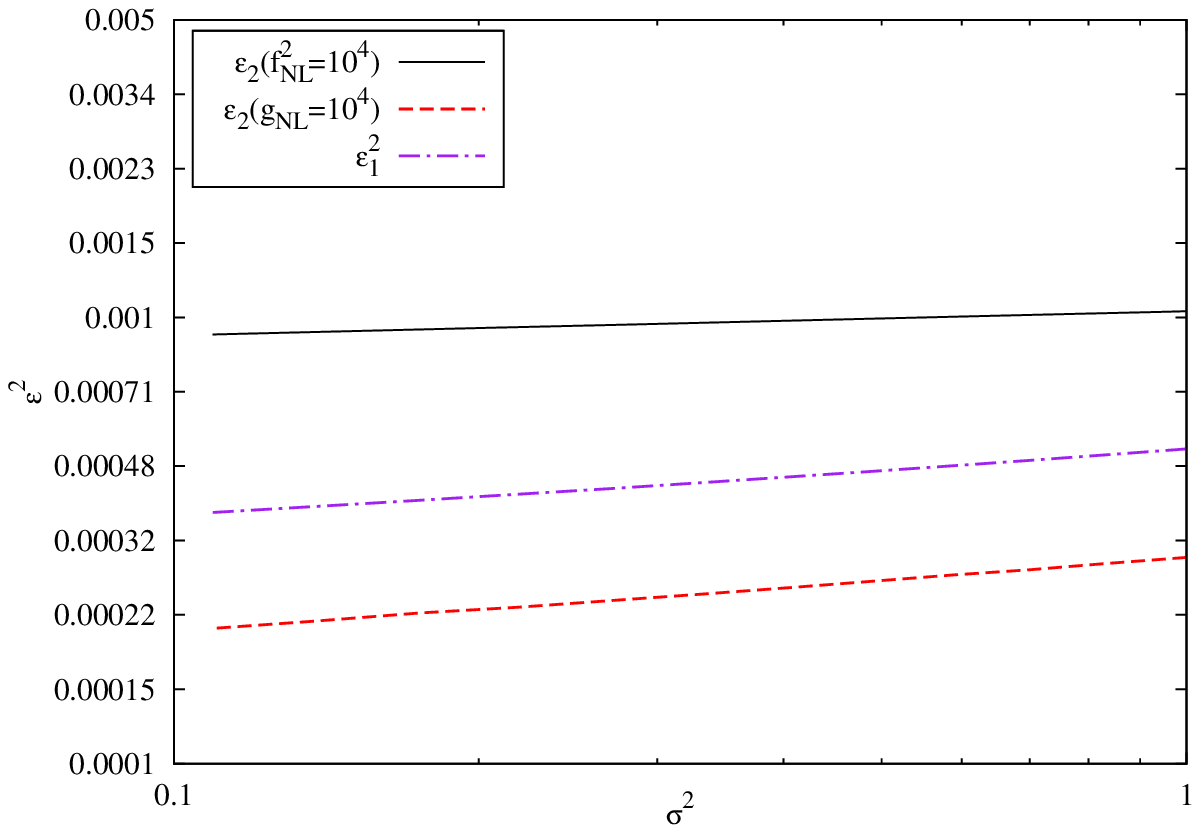}}
\la{fig-eps-b}
\caption{\footnotesize Scale dependence of the $\eps_n$. Panel
  (a) : Behaviour of $\eps_1$ vs. $\sig^2$ in the local and
  equilateral models, for $\fnl=100$ in each case. Panel (b) : 
  Behaviour of $\eps_2$ for the local model with $\fnl=100$ and
  $g_{\rm NL}=10^4$. The terms proportional to $\fnl^2$ and \gnl\ are
  shown separately. Also shown is $\eps_1^2$ for the same model. The
  axes are logscale.}
\la{fig-eps}
\end{figure}
While this choice allows us to have a well-defined relation between
length scales and masses, namely $M=(4\pi/3)\Om_m\rho_cR^3$ with 
$\rho_c = 3 H_0^2/(8 \pi G) = 2.77\cdot10^{11}
h^{-1}\Msol(h^{-1}{\rm Mpc})^{-3}$, it spoils the Markovianity of the
random walk of $\del_R$ (see e.g. \Cite{Maggiore:2009rv}). We will
therefore continue to present results for the sharp-k filter. By using
\eqns{delR} and \eqref{transfer} we have, for the $3$-point function, 
\be
\avg{\del_{R_1} \del_{R_2} \del_{R_3}}_c
= \int \frac{d^3 k_1}{(2 \pi)^3} \frac{d^3 k_2}{(2 \pi)^3} \frac{d^3
  k_3}{(2 \pi)^3} \tW(k_1 R_1) \tW(k_2 R_2) \tW(k_3 R_3) \Cal{M}(k_1)
\Cal{M}(k_2) \Cal{M}(k_3) \avg{\R(\k_1) \R(\k_2) \R(\k_3)}_c \, ,
\ee
where we suppressed the redshift dependence, and analogous
formulae are valid for the higher order correlations. 
\fig{fig-eps} shows the behaviour of the NG functions $\eps_1$ and
$\eps_2$ defined in the text (\eqn{eps-n}) as a function of
$S=\sig^2_R$ in the local and equilateral cases. ($\eps_2$ is shown
only for the local case.) We see that these functions remain
approximately constant over a range of $S$ values which corresponds to
roughly the range $2$-$25h^{-1}$Mpc of Lagrangian smoothing length
scales.

\section{\normalsize Some results concerning statistics of random
  walks}
\la{app-randomwalkstats}
\subsection{\normalsize Expressing \Fsw\ as a sum of Gaussians}
\la{app-FsvdW-resum}
\noindent
Let \Cal{F}\ denote the r.h.s of \eqn{FsvdW}. To prove the equivalence
of the two expressions for \Fsw\ in \eqns{FsvdW-inf} and
\eqref{FsvdW}, we will show that the Laplace transform
$\mathcal{L}(\omega)$ of \Cal{F}\ is the SvdW result \eqref{LsvdW}.
We have
\begin{align}
  \mathcal{L}(\omega)
  &= \int_0^{+\infty}\!\!\!\mathrm{d}S \, e^{-\omega S} \mathcal{F}(S)
  \nonumber\\
  &=\sum_{j=-\infty}^{+\infty} \frac{\delv-2j\delT}{(2\pi)^{1/2}}
  \int_0^{+\infty} \!\!\mathrm{d}S\, \frac{1}{S^{3/2}} e^{-\omega S
    -(\delv-2j\delT)^2/2S} \nonumber\\
  &= - \sum_{j=-\infty}^{+\infty} \mathrm{sgn} (2j\delT-\delv) \,
  e^{-|2j\delT-\delv|\sqrt{2\omega}} \notag \\
  &= e^{-\delv\sqrt{2\omega}} \sum_{j\leq 0} e^{2j\delT\sqrt{2\omega}}
  - e^{\delv\sqrt{2\omega}} \sum_{j>0} e^{-2j\delT\sqrt{2\omega}}\,,
\la{Fsvdw-Laptransf-1}
\end{align}
where the integral in the second line can be calculated using
Eqn. 3.472(5) of \Cite{GradRyzh}, and we used the fact
that $2j\delT-\delv$ is negative for $j\leq 0$. By changing
$j\rightarrow -j$ in the first summation in the last line one gets 
\begin{align}
  \mathcal{L}(\omega)
  &= e^{-\delv\sqrt{2\omega}} \sum_{j=0}^{+\infty}
  \left(e^{-2\delT\sqrt{2\omega}}\right)^j
  - e^{\delv\sqrt{2\omega}} \sum_{j=1}^{+\infty}
  \left(e^{-2\delT\sqrt{2\omega}}\right)^j \notag \\
  &=\frac{e^{-\delv\sqrt{2\omega}} - e^{\delv\sqrt{2\omega}}e^{-2\delT\sqrt{2\omega}}}
  {1 - e^{-2\delT\sqrt{2\omega}}}
  = \frac{e^{\delc\sqrt{2\omega}} - e^{-\delc\sqrt{2\omega}}}
  {e^{\delT\sqrt{2\omega}} - e^{-\delT\sqrt{2\omega}}}\,,
\la{Fsvdw-Laptransf-2}
\end{align}
which reproduces exactly the result of \eqn{LsvdW}.

\subsection{\normalsize Wrong side counting for the single barrier}
\la{app-wrongsidecounting}
\noindent
Here we show that the wrong side (WS) counting argument for the single 
barrier, is equivalent to MR's calculation of the f.c. rate
for generic random walks. For a single barrier problem with discrete
time steps, the probability for the walk to remain below the barrier
\delc\ for the first $n-1$ steps, and to cross the barrier at the
$n^{\rm th}$ step is $\int_{\delc}^\infty d\del_n
\int_{-\infty}^{\delc} d\del_{n-1} \ldots d\del_1
W(\del_0;\{\del_k\}_n;S_n)$. Here $W(\del_0;\{\del_k\}_n;S_n)$ 
defined in \eqn{W-def} is the p.d.f. of a completely generic random
walk -- we  are \emph{not} assuming that the walk is Markovian or even
Gaussian. Using an argument very similar to the one used in writing
\eqn{FSvdW-pathinteg}, the WS f.c. rate is then given by
\be
\Cal{F}_{\rm WS, 1-bar}(S) = \lim_{\DS\to0} \frac{1}{\DS}
\int_{\delc}^\infty d\del_n \int_{-\infty}^{\delc} d\del_{n-1} \ldots
d\del_1 W(\del_0;\{\del_k\}_n;S_n=S) \,.
\la{app-fcrate-WS}
\ee
In general we also have the identity
\be
\int_{-\infty}^\infty d\del_n \int_{-\infty}^{\delc} d\del_{n-1}
\ldots d\del_1 W(\del_0;\{\del_k\}_n;S_n) = \int_{-\infty}^{\delc}
d\del_{n-1} \ldots d\del_1 W(\del_0;\{\del_k\}_{n-1};S_{n-1}) \,,
\la{app-fcrate-identity}
\ee
which is a simple marginalisation over the location of the walk at the
final time step, and follows from the definition of
$W(\del_0;\{\del_k\}_n;S_n)$. Let $P_{\DS}(>S_n)$ denote the
probability that the f.c. time is larger than $S_n$. Since this is the
same as the probability that the walk has not crossed the barrier in
the first $n$ steps, we have
\be
P_{\DS}(>S_n) = \int_{-\infty}^{\delc}
d\del_{n} \ldots d\del_1 W(\del_0;\{\del_k\}_{n};S_{n})\,,
\la{app-fc-cumul}
\ee
and the r.h.s. of \eqn{app-fcrate-identity} equals
$P_{\DS}(>S_{n-1})$. Splitting the integral over $\del_n$ on the
l.h.s. of \eqn{app-fcrate-identity} as $\int_{-\infty}^{\infty}\to
\int_{-\infty}^{\delc}+ \int_{\delc}^{\infty}$ and using
\eqn{app-fc-cumul} we then get
\be
\int_{\delc}^\infty d\del_n \int_{-\infty}^{\delc} d\del_{n-1} 
\ldots d\del_1 W(\del_0;\{\del_k\}_n;S_n) = P_{\DS}(>S_{n-1}) -
P_{\DS}(>S_n) \,.
\la{app-fcrate-interm}
\ee
From \eqn{app-fcrate-WS} it then follows that
\be
\Cal{F}_{\rm WS, 1-bar}(S) = \lim_{\DS\to0} \frac{1}{\DS} \left[
  P_{\DS}(>S_{n-1}) - P_{\DS}(>S_n)\right] = -\p_S P_{\DS=0}(>S) =
\Cal{F}_{\rm MR, 1-bar}(S)\,,
\la{app-fcrate-final}
\ee
where $P_{\DS=0}(>S)$ is the continuum limit cumulative probability
for the f.c. time, which is what MR use to compute the f.c. rate. This
proves our result.

\subsection{\normalsize Comparison with Lam \etal\ \cite{Lam:2009nd}}
\la{app-comparelam}
\noindent
To compare with the analysis of Lam \etal\ \cite{Lam:2009nd} it is
useful to express the two barrier conditional f.c. rate in a slightly
different form. First, by manipulating the integrals in its
definition, the conditional probability density
$\Pi_{\DS}(\del_0,\del_n;S)$ can be rewritten as 
\begin{align}
\Pi_{\DS}(\del_0,\del_n;S_n)  &\equiv \int_{-\delv}^{\delc} d\del_{n-1}
\ldots d\del_1 W(\del_0;\{\del_k\}_n;S_n) \nonumber\\ 
&= \int_{-\delv}^{\infty}d\del_{n-1}\ldots
d\del_1 W(\del_0;\{\del_k\}_n;S_n) \nonumber\\
&\ph{\int_{-\delv}^{\infty}d\del_{n-1}}
- \sum_{k=1}^{n-1}
\int_{-\delv}^{\infty}d\del_{n-1}\ldots d\del_{k+1} 
\int_{\delc}^{\infty}d\del_k
\int_{-\delv}^{\delc} d\del_{k-1} \ldots d\del_1
W(\del_0;\{\del_k\}_n;S_n)\,,
\la{compare-lam-1}
\end{align}
where it is understood that for $k=1$ the integral in the second term
begins with $\int_{\delc}^{\infty} d\del_1$ on the far right, and for
$k=n-1$ it ends with $\int_{\delc}^{\infty} d\del_{n-1}$ on the far
left. This form of the expression is useful because it helps us to
intuitively understand the calculation of the conditional f.c. rate,
which becomes  
\begin{align}
\Fng(S) &\equiv \lim_{\DS\to0}\frac{1}{\DS}
\int_{-\infty}^{-\delv}d\del_n \Pi_{\DS}(\del_0,\del_n;S) \nonumber\\ 
&=\lim_{\DS\to0}\frac{1}{\DS} \int_{-\infty}^{-\delv}d\del_n
\int_{-\delv}^{\infty}d\del_{n-1}\ldots d\del_1
W(\del_0;\{\del_k\}_n;S)  \nonumber\\
&\ph{\lim_{\DS\to0}\frac{1}{\DS}}
- \lim_{\DS\to0}\frac{1}{{\DS}}\sum_{k=1}^{n-1} 
 \int_{-\infty}^{-\delv}d\del_n
\int_{-\delv}^{\infty}d\del_{n-1}\ldots d\del_{k+1}  
\int_{\delc}^{\infty}d\del_k
\int_{-\delv}^{\delc} d\del_{k-1} \ldots d\del_1
W(\del_0;\{\del_k\}_n;S)\,.
\la{compare-lam-2}
\end{align}
The first term here is simply the single barrier rate across the void
barrier (see Appendix \ref{app-wrongsidecounting}). The second term
subtracts from this rate the fraction of trajectories which crossed
the halo barrier at times before $S$, without having crossed the void
barrier before. This is exactly what Lam \etal\ write in
Eqn. 41 of \Cite{Lam:2009nd} (in the continuum limit the summation
over $k$ above would become an integral over time). Their expression
however assumes a factorisation of the second term into a two barrier
rate convolved with a conditional probability. This is certainly true
for the Gaussian case with the sharp-k filter, as one can immediately
see by replacing $W$ above with $W^{\rm gm}$ which factorises, giving
the two barrier probability across the \emph{halo} barrier, convolved
with the usual Gaussian conditional probability. In the non-Gaussian
case however, $W(\del_0;\{\del_k\}_n;S)$ does not factorise due to the
presence of multi-scale correlations which make the process
non-Markovian. In practice though these effects are small, and hence
factorisability should be a reasonable assumption. Notice however that
our approach does not require this assumption, and we also do not need
to linearize the effects of non-Gaussianities as in \Cite{Lam:2009nd}.

\section{\normalsize Details of the calculation of \Fsw}
\la{app-fsvdw-details}
\subsection{\normalsize $\Cal{I}^{\DS}_n(\del)$ for fixed
  $\del<-\delv$ and small \DS}
\la{app-fsvdw-details-1}
\noindent
With $\del<-\delv$ held fixed, a few manipulations allow us to write
$\Cal{I}^{\DS}_n(\del)$ as
\begin{align}
\Cal{I}^{\DS}_n(\del<-\delv)) &= \frac{(2\DS)^{n/2}}{\pi^{1/2}} \left[ 
  \int_{-\infty}^{(\del+\delv)/\sqrt{2\DS}} dy \,y^n e^{-y^2} -
  \int_{-\infty}^{(\del-\delc)/\sqrt{2\DS}} dy \,y^n e^{-y^2} \right]
\nonumber\\ 
&= \frac{(2\DS)^{n/2}}{\pi^{1/2}}(-1)^n \left[
  \int_{|\del+\delv|/\sqrt{2\DS}}^{\infty} dq \,q^n e^{-q^2} - \ldots
  \right] \,,
\la{app-IDelSn-1}
\end{align}
where we ignore the second term since one can explicitly check that it
leads to an exponentially suppressed contribution which eventually
vanishes in the appropriate continuum limit. (This essentially follows
from the fact that $\delT=\delc+\delv$ is finite and hence
$\delT/\sqrt{2\DS}\to+\infty$ as $\DS\to0$.)
Now, since the variable $q$ in the last integral is strictly positive
under our assumptions, one can check that the transformation to
$x=q^2$ will reduce the integral above to an incomplete Gamma
function, 
\be
\Cal{I}^{\DS}_n(\del<-\delv) = \frac{(2\DS)^{n/2}}{\pi^{1/2}}(-1)^n
\left[ \frac{1}{2}\Gamma\left(
  \frac{n+1}{2},\frac{(\del+\delv)^2}{2\DS}\right) +\ldots \right]  \,,
\la{app-IDelSn-2}
\ee
and in the limit of small \DS, the asymptotic expansion of the
incomplete Gamma gives us \eqn{Inofdel-lessdelv}.

\section{\normalsize Details of the calculation of \Fng}
\la{app-fng-details}
\subsection{\normalsize The two barrier derivative exchange property}
\la{app-fng-details-1}
\noindent
To prove \eqn{deriv-exch-2bar}, for any function $g(\del_1,\ldots
,\del_n)$ define the function $h(\del_n;\delc,\delv)$  of three
variables as the multiple integral 
\be
h(\del_n;\delc,\delv) = \int_{-\delv}^{\delc}d\del_{n-1}\ldots d\del_1
\, g(\del_1,\ldots ,\del_n)\,.
\la{app-h-def}
\ee
The l.h.s. of \eqn{deriv-exch-2bar} can then be reduced as
\begin{align}
\textrm{L.h.s.} &=
\int_{-\infty}^{-\delv}d\del_n\int_{-\delv}^{\delc}d\del_{n-1}\ldots
d\del_1 \sum_{j=1}^n{\p_j}g \nonumber\\
&= h(-\delv;\delc,\delv) + \sum_{j=1}^{n-1}
\int_{-\infty}^{-\delv}d\del_n\int_{-\delv}^{\delc}d\del_{n-1}\ldots
d\del_1 {\p_j}g \nonumber\\
&= h(-\delv;\delc,\delv) + \int_{-\infty}^{-\delv}d\del_n \left(
\p_{\delc}|_{\delv} - \p_{\delv}|_{\delc}\right)
h(\del_n;\delc,\delv)\,, 
\la{app-derivexch-lhs}
\end{align}
where in the second line we integrated the term involving
$\p_{\del_n}$ and the third line follows since for each $1\leq
j\leq n-1$, the integral becomes
\be
\int_{-\infty}^{-\delv}d\del_n\int_{-\delv}^{\delc}d\del_{n-1}\ldots
d\del_{j+1}d\del_{j-1}\ldots d\del_1 \bigg[g(\del_1,\ldots,
\del_j=\delc,\ldots,\del_n) - g(\del_1,\ldots,
\del_j=-\delv,\ldots,\del_n) \bigg]\,,
\la{app-derivexch-interm-1}
\ee
the summation of which is the same as the integral in the third
line. The derivative $\p_{\delc}|_{\delv}$ now simply comes across the
integral over $\del_n$ since the boundary is independent of \delc. The
term involving $\p_{\delv}|_{\delc}$ can be handled by noting that
when this derivative acts only on the boundary of the integral
$\int_{-\infty}^{-\delv} d\del_n \,h(\del_n;\delc,\delv)$, we simply
get $-h(-\delv;\delc,\delv)$. We can then write
\be
\p_{\delv}|_{\delc} \int_{-\infty}^{-\delv} d\del_n\,
h(\del_n;\delc,\delv) = -h(-\delv;\delc,\delv) +
\int_{-\infty}^{-\delv} d\del_n\, \p_{\delv}|_{\delc}
h(\del_n;\delc,\delv)\,.
\la{app-derivexch-interm-2}
\ee
This simplifies the expression \eqref{app-derivexch-lhs} to 
\be
\textrm{L.h.s.} = \left( \p_{\delc}|_{\delv}
-\p_{\delv}|_{\delc}\right) \int_{-\infty}^{-\delv}d\del_n\,
h(\del_n;\delc,\delv)\,.
\la{app-derivexch-lhs-2}
\ee
It is also straightforward to show that under the change of variables
$(\delc,\delv)\to (\delT,\delv)$ where $\delT=\delc+\delv$, we have
$\p_{\delc}|_{\delv} -\p_{\delv}|_{\delc} = -\p_{\delv}|_{\delT}$, which
gives us
\be
\textrm{L.h.s.} =  -\left. \frac{\p}{\p\delv}\right|_{\delT}
\int_{-\infty}^{-\delv}d\del_n \int_{-\delv}^{\delc} d\del_{n-1}\ldots
d\del_1\,g(\del_1,\ldots, \del_n) = \textrm{R.h.s.}\,,
\la{app-derivexch-rhs}
\ee
which proves the result.

\subsection{\normalsize Leading unequal time contribution}
\la{app-fng-details-2}
\noindent
Here we sketch a proof of \eqn{F3NL-eval}. Defining the quantity $M_j$
as
\be
M_j \equiv \sum_{k,l=1}^n\int_{-\infty}^{-\delv}d\del_n
\int_{-\delv}^{\delc}d\del_{n-1} \ldots d\del_1 \p_k\p_l \left( \p_j
W^{\rm gm}\right)\,,
\la{app-Mj-def}
\ee
we can write
\be
\Cal{F}^{(\rm 3,NL)} = \lim_{\DS\to 0}\frac{1}{2\DS}
\Cal{G}_3^{(1,0,0)}(S) \sum_j(S-S_j) M_j\,.
\la{app-F3NL}
\ee
Using the derivative exchange property \eqref{deriv-exch-2bar} we can
write $M_j$ as
\begin{align}
M_j &= \p_{\delv}^2\int_{-\infty}^{-\delv}d\del_n
\int_{-\delv}^{\delc}d\del_{n-1} \ldots d\del_1\, \p_j W^{\rm gm}
\nonumber\\ 
& = -\p_{\delv}^2\int_{-\infty}^{-\delv}d\del_n \left[ \Pi^{\rm
    gm}_{\DS}(\del_0,-\delv;S_j)\Pi^{\rm gm}_{\DS}(-\delv,\del_n;S-S_j)
  - \Pi^{\rm gm}_{\DS}(\del_0,\delc;S_j)\Pi^{\rm
    gm}_{\DS}(\delc,\del_n;S-S_j) \right] \,.
\la{app-Mj-interim1}
\end{align}
Recall that $\p_{\delv} = \p/\p\delv|_{\delT}$. In the second line, we
already have an expression for $\Pi^{\rm gm}_{\DS}(\del_0,-\delv;S_j)$
from \eqn{Pi-gm-arbit>explicit}. One can also check that in the
continuum limit, the second term involving \delc\ will be
exponentially suppressed. The only non-trivial quantity we need then
is the integral $\int_{-\infty}^{-\delv}d\del_n\Pi^{\rm
  gm}_{\DS}(-\delv,\del_n;S-S_j)$. Following techniques similar to
those discussed in Section \ref{FSvdW-PIdetails} leading up to
\eqn{Pi-gm-WSinteg}, we can write this integral by simply replacing
$\del_0$ in \eqref{Pi-gm-WSinteg} with $(-\delv)$ to get
\be
\int_{-\infty}^{-\delv}d\del_n\Pi^{\rm gm}_{\DS}(-\delv,\del_n;S-S_j)
= \left(\frac{\DS}{2\pi}\right)^{1/2} \left[
  \int_{-\infty}^0\frac{d\eta}{(-\eta)}v(\eta)e^{-\eta^2} \right]
\Pi^{\rm gm}_{\DS}(-\delv,-\delv;S-S_j)\,,
\la{app-Mj-interim2}
\ee
with the same boundary layer function $v(\eta)$ appearing, since this
is independent of the value of $\del_0$. Notice that the integral
involving $v(\eta)$ is identical to that appearing in the expression
\eqref{FSvdW-derived} for \Fsw. In fact the other constant $\gamma$ in
that expression also appears in $M_j$ through the object $\Pi^{\rm
    gm}_{\DS}(\del_0,-\delv;S_j)$, and will lead us to identify a
factor of \Fsw\ in the final expression, see below. The object
$\Pi^{\rm gm}_{\DS}(-\delv,-\delv;S-S_j)$ is similar to one discussed
by MR in \Cite{Maggiore:2009rv}, where they showed that $\Pi^{\rm 
  gm}_{1-bar,\DS}(\delc,\delc;\ti{S}) = \DS/(2\pi \ti{S}^3)^{1/2}$,
independent of \delc. In fact, one can show that $\Pi^{\rm
  gm}_{\DS}(-\delv,-\delv;\ti{S})$ is also independent of \emph{both}
\delc\ and \delv, and has the same value  
\be
\Pi^{\rm gm}_{\DS}(-\delv,-\delv;\ti{S}) =\frac{\DS}{\sqrt{2\pi}}
\frac{1}{\ti{S}^{3/2}}\,.
\la{app-Mj-interim3}
\ee
This can be checked in two steps. Firstly for arbitrary $n$ one can
show that the path integral with $n$ steps involved in $\Pi^{\rm
  gm}_{\DS}(-\delv,-\delv;S_n)$ becomes independent of \delc\ and
\delv\ in the continuum limit (since \delc\ and \delv\ only appear in
the limits of integration, and a suitable change of variables sends
these to infinity as $\DS\to0$). The remaining integral is therefore
only a function of $n$. Following MR, a dimensional argument coupled
with a straightforward calculation for $n=2$ then fixes the time
dependence completely, leading to \eqn{app-Mj-interim3}.
Putting things together, 
\begin{align}
M_j &= -\p_{\delv}^2 \left[ \Pi^{\rm
    gm}_{\DS}(\del_0,-\delv;S_j)\left(
  \left(\frac{\DS}{2\pi}\right)^{1/2} \left[
    \int_{-\infty}^0\frac{d\eta}{(-\eta)}v(\eta)e^{-\eta^2} \right] 
\right) \left( \frac{\DS}{\sqrt{2\pi}} \frac{1}{(S-S_j)^{3/2}} \right)
\right]  \nonumber\\ 
&=-\frac{(\DS)^2}{\sqrt{2\pi}}\p_{\delv}^2\,
\frac{\Fsw(S_j)}{(S-S_j)^{3/2}} \,, 
\la{app-Mj-final}
\end{align}
where we used \eqns{Pi-gm-arbit>explicit} and \eqn{FSvdW-derived} in
writing the last line. Using this result for $M_j$ in \eqn{app-F3NL}
with the continuum limit $\sum_{j=1}^nh(S_j) \to\int_0^S d\ti{S}
h(\ti{S})(\DS)^{-1}$, we see that the factors of \DS\ cancel, and
further expressing $\Cal{G}_3^{(1,0,0)}(S)$ in terms of $c_1(S)$
finally leads to the result of \eqn{F3NL-eval}.

{

}

\begin{thebibliography}{99}
\bibitem{Kirshner:1981wz}
  R.~P.~Kirshner, A.~.~J.~Oemler, P.~L.~Schechter and S.~A.~Shectman,
  ``A million cubic megaparsec void in Bootes,''
  Astrophys.\ J.\  {\bf 248}, L57 (1981).

\bibitem{Patiri:2005ys}
  S.~G.~Patiri, J.~Betancort-Rijo, F.~Prada, A.~Klypin and S.~Gottlober,
  ``Statistics of Voids in the 2dF Galaxy Redshift Survey,''
  Mon.\ Not.\ Roy.\ Astron.\ Soc.\  {\bf 369}, 335 (2006)
  [arXiv:astro-ph/0506668].

\bibitem{Goldberg:2004jr}
  D.~M.~Goldberg, T.~D.~Jones, F.~Hoyle, R.~R.~Rojas, M.~S.~Vogeley and M.~R.~Blanton,
  ``The Mass Function of Void Galaxies in the SDSS Data Release 2,''
  Astrophys.\ J.\  {\bf 621}, 643 (2005)
  [arXiv:astro-ph/0406527].

\bibitem{Croton:2004ac}
  D.~J.~Croton {\it et al.}  [The 2dFGRS Team Collaboration],
  ``The 2dF Galaxy Redshift Survey: Voids and hierarchical scaling models,''
  Mon.\ Not.\ Roy.\ Astron.\ Soc.\  {\bf 352}, 828 (2004)
  [arXiv:astro-ph/0401406].

\bibitem{Hoyle:2001kn}
  F.~Hoyle and M.~S.~Vogeley,
  ``Voids in the PSCz Survey and the Updated Zwicky Catalog,''
  Astrophys.\ J.\  {\bf 566}, 641 (2002)
  [arXiv:astro-ph/0109357].

\bibitem{Hoyle:2005kn}
  F.~Hoyle, M.~S.~Vogeley and R.~R.~Rojas,
  ``Void Galaxies in the Sloan Digital SKy Survey,''
  Bulletin of the AAS\ {\bf 37}, 443 (2005)


\bibitem{ForeroRomero:2008ig}
  J.~E.~Forero-Romero, Y.~Hoffman, S.~Gottloeber, A.~Klypin and G.~Yepes,
  ``A Dynamical Classification of the Cosmic Web,''
  Mon.\ Not.\ Roy.\ Astron.\ Soc.\  {\bf 396}, 1815 (2009)
  [arXiv:0809.4135 [astro-ph]].

\bibitem{Lee:2008ar}
  J.~Lee and B.~Lee,
  ``The Variation of Galaxy Morphological Type with the Shear of Environment,''
  Astrophys.\ J.\ {\bf 688}, 78 (2008) 
  [arXiv:0801.1558 [astro-ph]].

\bibitem{AragonCalvo:2007mk}
  M.~A.~Aragon-Calvo, B.~J.~T.~Jones, R.~van de Weygaert and M.~J.~van der Hulst,
  ``The Multiscale Morphology Filter: Identifying and Extracting Spatial
  Patterns in the Galaxy Distribution,''
  Astron.\ Astrophys.\  {\bf 474}, 315 (2004)
  [arXiv:0705.2072 [astro-ph]].

\bibitem{Shandarin:2009ue}
  S.~Shandarin, S.~Habib and K.~Heitmann,
  ``Origin of the Cosmic Network: Nature vs Nurture,''
  Phys.\ Rev.\  D {\bf 81}, 103006 (2010)
  [arXiv:0912.4471 [astro-ph.CO]].


\bibitem{vandeWeygaert:2009vy}
  R.~van de Weygaert, M.~A.~Aragon-Calvo, B.~J.~T.~Jones and E.~Platen,
  ``Geometry and Morphology of the Cosmic Web: Analyzing Spatial Patterns in
  the Universe,''
  arXiv:0912.3448 [astro-ph.IM].

\bibitem{Peacock:2001gs}
  J.~A.~Peacock {\it et al.},
  ``A measurement of the cosmological mass density from clustering in the 2dF
  Galaxy Redshift Survey,''
  Nature {\bf 410}, 169 (2001)
  [arXiv:astro-ph/0103143].

\bibitem{Eisenstein:2005su}
  D.~J.~Eisenstein {\it et al.}  [SDSS Collaboration],
  ``Detection of the Baryon Acoustic Peak in the Large-Scale Correlation
  Function of SDSS Luminous Red Galaxies,''
  Astrophys.\ J.\  {\bf 633}, 560 (2005)
  [arXiv:astro-ph/0501171].

\bibitem{Abazajian:2008wr}
  K.~N.~Abazajian {\it et al.}  [SDSS Collaboration],
  ``The Seventh Data Release of the Sloan Digital Sky Survey,''
  Astrophys.\ J.\ Suppl.\  {\bf 182}, 543 (2009)
  [arXiv:0812.0649 [astro-ph]].

\bibitem{Cimatti:2009is}
  A.~Cimatti {\it et al.},
  ``Euclid Assessment Study Report for the ESA Cosmic Visions,''
  arXiv:0912.0914 [astro-ph.CO].

\bibitem{Cappelluti:2010ay}
  N.~Cappelluti {\it et al.},
  ``eROSITA on SRG: a X-ray all-sky survey mission,''
  arXiv:1004.5219 [astro-ph.IM].

\bibitem{Schlegel:2009hj}
  D.~Schlegel, M.~White and D.~Eisenstein  [with input from the SDSS-III
                  collaboration and with input from the SDSS-III],
  ``The Baryon Oscillation Spectroscopic Survey: Precision measurements of the
  absolute cosmic distance scale,''
  arXiv:0902.4680 [astro-ph.CO].

\bibitem{Maldacena:2002vr}
  J.~M.~Maldacena,
  ``Non-Gaussian features of primordial fluctuations in single field
  inflationary models,''
  JHEP {\bf 0305}, 013 (2003)
  [arXiv:astro-ph/0210603].

\bibitem{Acquaviva:2002ud}
  V.~Acquaviva, N.~Bartolo, S.~Matarrese and A.~Riotto,
  ``Second-order cosmological perturbations from inflation,''
  Nucl.\ Phys.\  B {\bf 667}, 119 (2003)
  [arXiv:astro-ph/0209156].

\bibitem{Komatsu:2010fb}
  E.~Komatsu {\it et al.},
  ``Seven-Year Wilkinson Microwave Anisotropy Probe (WMAP) Observations: Cosmological Interpretation,''
  arXiv:1001.4538 [astro-ph.CO].

\bibitem{Slosar:2008hx}
  A.~Slosar, C.~Hirata, U.~Seljak, S.~Ho and N.~Padmanabhan,
  ``Constraints on local primordial non-Gaussianity from large scale
  structure,''
  JCAP {\bf 0808} (2008) 031
  [arXiv:0805.3580 [astro-ph]].


\bibitem{Sartoris:2010cr}
  B.~Sartoris, S.~Borgani, C.~Fedeli, S.~Matarrese, L.~Moscardini, P.~Rosati and J.~Weller,
  ``The potential of X-ray cluster surveys to constrain primordial non-Gaussianity,''
  arXiv:1003.0841 [astro-ph.CO].

\bibitem{Cunha:2010zz}
  C.~Cunha, D.~Huterer and O.~Dore,
  ``Primordial non-Gaussianity from the covariance of galaxy cluster counts,''
  Phys.\ Rev.\  D {\bf 82}, 023004 (2010)
  [arXiv:1003.2416 [astro-ph.CO]].

\bibitem{Dalal:2007cu}
  N.~Dalal, O.~Dore, D.~Huterer and A.~Shirokov,
  ``The imprints of primordial non-Gaussianities on large-scale
  structure: scale dependent bias and abundance of virialized
  objects,'' 
  Phys.\ Rev.\  D {\bf 77} (2008) 123514
  [arXiv:0710.4560 [astro-ph]].

\bibitem{Matarrese:2008nc}
  S.~Matarrese and L.~Verde,
  ``The effect of primordial non-Gaussianity on halo bias,''
  Astrophys.\ J.\  {\bf 677} (2008) L77
  [arXiv:0801.4826 [astro-ph]].


\bibitem{Sefusatti:2009qh}
  E.~Sefusatti,
  ``1-loop Perturbative Corrections to the Matter and Galaxy Bispectrum with
  non-Gaussian Initial Conditions,''
  Phys.\ Rev.\  D {\bf 80} (2009) 123002
  [arXiv:0905.0717 [astro-ph.CO]].

\bibitem{Matarrese:2000iz}
  S.~Matarrese, L.~Verde and R.~Jimenez,
  ``The abundance of high-redshift objects as a probe of non-Gaussian initial conditions,''
  Astrophys.\ J.\  {\bf 541} (2000) 10
  [arXiv:astro-ph/0001366].

\bibitem{LoVerde:2007ri}
  M.~LoVerde, A.~Miller, S.~Shandera and L.~Verde,
  ``Effects of Scale-Dependent Non-Gaussianity on Cosmological Structures,''
  JCAP {\bf 0804} (2008) 014
  [arXiv:0711.4126 [astro-ph]].

\bibitem{Verde:2010wp}
  L.~Verde,
  ``Non-Gaussianity from Large-Scale Structure Surveys,''
  arXiv:1001.5217 [astro-ph.CO].

\bibitem{Desjacques:2010jw}
  V.~Desjacques and U.~Seljak,
  ``Primordial non-Gaussianity from the large scale structure,''
  Class.\ Quant.\ Grav.\  {\bf 27}, 124011 (2010)
  [arXiv:1003.5020 [astro-ph.CO]].

\bibitem{Colberg:2008qg}
  J.~M.~Colberg {\it et al.},
  ``The Aspen--Amsterdam Void Finder Comparison Project,''
  Mon.\ Not.\ Roy.\ Astron.\ Soc.\  {\bf 387}, 933 (2008)
  [arXiv:0803.0918 [astro-ph]].

\bibitem{vandeWeygaert:2009hr}
  R.~van de Weygaert and E.~Platen,
  ``Cosmic Voids: structure, dynamics and galaxies,''
  arXiv:0912.2997 [astro-ph.CO].

\bibitem{Dubinski:1992tr}
  J.~Dubinski, L.~Nicolaci da Costa, D.~S.~Goldwirth, M.~Lecar and T.~Piran,
  ``Void evolution and the large scale structure,''
  Astrophys.\ J.\  {\bf 410}, 458 (1993).

\bibitem{Colberg:2004nd}
  J.~M.~Colberg, R.~K.~Sheth, A.~Diaferio, L.~Gao and N.~Yoshida,
  Mon.\ Not.\ Roy.\ Astron.\ Soc.\  {\bf 360}, 216 (2005)
  [arXiv:astro-ph/0409162].

\bibitem{vandeWeygaert:1993cp}
  R.~van de Weygaert and E.~van Kampen,
  ``Voids in Gravitational Instability Scenarios - Part One - Global Density and Velocity Fields in an Einstein - De-Sitter Universe,''
  Mon.\ Not.\ Roy.\ Astron.\ Soc.\  {\bf 263}, 481 (1993).

\bibitem{Blumenthal:1992tr}
  G.~R.~Blumenthal, L.~Nicolaci da Costa, D.~S.~Goldwirth, M.~Lecar and T.~Piran,
  ``The largest possible voids,''
  Astrophys.\ J.\  {\bf 388}, 234 (1992).

\bibitem{Sheth:2003py}
  R.~K.~Sheth and R.~van de Weygaert,
  ``A hierarchy of voids: Much ado about nothing,''
  Mon.\ Not.\ Roy.\ Astron.\ Soc.\  {\bf 350}, 517 (2004)
  [arXiv:astro-ph/0311260].

\bibitem{Press:1973iz}
  W.~H.~Press and P.~Schechter,
  ``Formation of galaxies and clusters of galaxies by selfsimilar gravitational
  condensation,''
  Astrophys.\ J.\  {\bf 187}, 425 (1974).

\bibitem{Epstein:1983cp}
  R.~I.~Epstein,
  ``Proto-galactic perturbations,''
  Mon.\ Not.\ Roy.\ Astron.\ Soc.\  {\bf 205}, 207 (1983).

\bibitem{Bond:1990iw}
  J.~R.~Bond, S.~Cole, G.~Efstathiou and N.~Kaiser,
  ``Excursion set mass functions for hierarchical Gaussian fluctuations,''
  Astrophys.\ J.\  {\bf 379}, 440 (1991).


\bibitem{Gunn:1972sv}
  J.~E.~Gunn and J.~R.~I.~Gott,
  ``On the infall of matter into cluster of galaxies and some effects on their evolution,''
  Astrophys.\ J.\  {\bf 176}, 1 (1972).

\bibitem{Fillmore:1984nj}
 J.~A.~Fillmore and P.~Goldreich,
  ``Self-similar spherical voids in an expanding universe,''
  Astrophys.\ J.\   {\bf 281}, 9 (1984).


\bibitem{Bertschinger:1985nj}
  E.~Bertschinger,
  ``The self-similar evolution of holes in an Einstein-de Sitter universe,''
  Astrophys.\ J.\ Suppl.\  {\bf 58}, 1 (1985).

\bibitem{Lavaux:2009wm}
  G.~Lavaux and B.~D.~Wandelt,
  ``Precision cosmology with voids: definition, methods, dynamics,''
  Mon.\ Not.\ Roy.\ Astron.\ Soc.\  {\bf 403}, 1392 (2010)
  [arXiv:0906.4101 [astro-ph.CO]].

\bibitem{Biswas:2010ey}
  R.~1.~Biswas, E.~Alizadeh and B.~D.~Wandelt,
  ``Voids as a Precision Probe of Dark Energy,''
  Phys.\ Rev.\  D {\bf 82}, 023002 (2010)
  [arXiv:1002.0014 [astro-ph.CO]].


\bibitem{Furlanetto:2005cc}
  S.~Furlanetto and T.~Piran,
  ``The Evidence of Absence: Galaxy Voids in the Excursion Set Formalism,''
  Mon.\ Not.\ Roy.\ Astron.\ Soc.\  {\bf 366}, 467 (2006)
  [arXiv:astro-ph/0509148].


\bibitem{Kamionkowski:2008sr}
  M.~Kamionkowski, L.~Verde and R.~Jimenez,
  ``The Void Abundance with Non-Gaussian Primordial Perturbations,''
  JCAP {\bf 0901}, 010 (2009)
  [arXiv:0809.0506 [astro-ph]].

\bibitem{Lam:2009nd}
  T.~Y.~Lam, R.~K.~Sheth and V.~Desjacques,
  ``The initial shear field in models with primordial local non-Gaussianity and
  implications for halo and void abundances,''
  arXiv:0905.1706 [astro-ph.CO].

\bibitem{Chongchitnan:2010xz}
  S.~Chongchitnan and J.~Silk,
  ``A Study of High-Order Non-Gaussianity with Applications to Massive Clusters
  and Large Voids,''
  Astrophys.\ J.\  {\bf 724}, 285 (2010)
  [arXiv:1007.1230 [astro-ph.CO]].

\bibitem{Maggiore:2009rx}
  M.~Maggiore and A.~Riotto,
  ``The halo mass function from the excursion set method. III. First
  principle derivation for non-Gaussian theories,'' 
  Astrophys.\ J.\  {\bf 717}, 526 (2010)
  [arXiv:0903.1251 [astro-ph.CO]].

\bibitem{D'Amico:2010ta}
  G.~D'Amico, M.~Musso, J.~Nore\~na and A.~Paranjape,
  ``An Improved Calculation of the Non-Gaussian Halo Mass Function,''
  arXiv:1005.1203 [astro-ph.CO].

\bibitem{Maggiore:2009rv}
  M.~Maggiore and A.~Riotto,
  ``The Halo Mass Function from the Excursion Set Method. I. First principle
  derivation for the non-Markovian case of Gaussian fluctuations and generic
  filter,''
  Astrophys.\ J.\  {\bf 711}, 907 (2010)
  [arXiv:0903.1249 [astro-ph.CO]].

\bibitem{Maggiore:2009rw}
  M.~Maggiore and A.~Riotto,
  ``The halo mass function from the excursion set method. II. The diffusing
  barrier,''
  Astrophys.\ J.\  {\bf 717}, 515 (2010)
  [arXiv:0903.1250 [astro-ph.CO]].

\bibitem{Hoffman:1984nj}
G.~L.~Hoffman, E.~E.~Salpeter and I.~Wasserman,
  ``Spherical simulations of holes and honeycombs in Friedmann universes,''
  Astrophys.\ J.\   {\bf 268}, 527 (1983).


\bibitem{Redner}
 S.~Redner, 
 ``A guide to first-passage processes,''
 Cambridge, UK : Cambridge Univ. Press (2001) 312pp.

\bibitem{D'Amico:future}
G. D'Amico \etal, in progress.

\bibitem{Sheth:1999su}
  R.~K.~Sheth, H.~J.~Mo and G.~Tormen,
  ``Ellipsoidal collapse and an improved model for the number and spatial
  distribution of dark matter haloes,''
  Mon.\ Not.\ Roy.\ Astron.\ Soc.\  {\bf 323}, 1 (2001)
  [arXiv:astro-ph/9907024].

\bibitem{Sheth:2001dp}
  R.~K.~Sheth and G.~Tormen,
  ``An Excursion Set Model Of Hierarchical Clustering : Ellipsoidal Collapse
  And The Moving Barrier,''
  Mon.\ Not.\ Roy.\ Astron.\ Soc.\  {\bf 329}, 61 (2002)
  [arXiv:astro-ph/0105113].

\bibitem{DeSimone:2010mu}
  A.~De Simone, M.~Maggiore and A.~Riotto,
  ``Excursion Set Theory for generic moving barriers and non-Gaussian initial
  conditions,''
  arXiv:1007.1903 [astro-ph.CO].

\bibitem{Lam:2009nb}
  T.~Y.~Lam and R.~K.~Sheth,
  ``Halo abundances in the \fnl model,''
  arXiv:0905.1702 [astro-ph.CO].

\bibitem{Babich:2004gb}
  D.~Babich, P.~Creminelli and M.~Zaldarriaga,
  ``The shape of non-Gaussianities,''
  JCAP {\bf 0408} (2004) 009
  [arXiv:astro-ph/0405356].

\bibitem{Lyth:2002my}
  D.~H.~Lyth, C.~Ungarelli and D.~Wands,
  ``The primordial density perturbation in the curvaton scenario,''
  Phys.\ Rev.\  D {\bf 67} (2003) 023503
  [arXiv:astro-ph/0208055].

\bibitem{Bartolo:2003jx}
  N.~Bartolo, S.~Matarrese and A.~Riotto,
  ``On non-Gaussianity in the curvaton scenario,''
  Phys.\ Rev.\  D {\bf 69} (2004) 043503
  [arXiv:hep-ph/0309033].

\bibitem{Dvali:2003em}
  G.~Dvali, A.~Gruzinov and M.~Zaldarriaga,
  ``A new mechanism for generating density perturbations from inflation,''
  Phys.\ Rev.\  D {\bf 69} (2004) 023505
  [arXiv:astro-ph/0303591].
  
\bibitem{Alishahiha:2004eh}
  M.~Alishahiha, E.~Silverstein and D.~Tong,
  ``DBI in the sky,''
  Phys.\ Rev.\  D {\bf 70}, 123505 (2004)
  [arXiv:hep-th/0404084].
  
\bibitem{ArkaniHamed:2003uz}
  N.~Arkani-Hamed, P.~Creminelli, S.~Mukohyama and M.~Zaldarriaga,
  ``Ghost Inflation,''
  JCAP {\bf 0404}, 001 (2004)
  [arXiv:hep-th/0312100].
    
\bibitem{Creminelli:2003iq}
  P.~Creminelli,
  ``On non-Gaussianities in single-field inflation,''
  JCAP {\bf 0310}, 003 (2003)
  [arXiv:astro-ph/0306122].

\bibitem{Creminelli:2005hu}
  P.~Creminelli, A.~Nicolis, L.~Senatore, M.~Tegmark and M.~Zaldarriaga,
  ``Limits on non-Gaussianities from WMAP data,''
  JCAP {\bf 0605} (2006) 004
  [arXiv:astro-ph/0509029].


\bibitem{Bardeen:1985tr}
  J.~M.~Bardeen, J.~R.~Bond, N.~Kaiser and A.~S.~Szalay,
  ``The Statistics Of Peaks Of Gaussian Random Fields,''
  Astrophys.\ J.\  {\bf 304}, 15 (1986).

\bibitem{GradRyzh}
I.~S.~Gradshteyn and I.~M.~Ryzhik,
``Tables of Integrals, Series and Products'',  7th ed.,
Amsterdam : Elsevier (2007) 1171pp. 

%
%
%
%
%
%
%
%
%
%
%
%
\end{thebibliography}
\end{document}